\documentclass[aps,prx,twocolumn,superscriptaddress,longbibliography]{revtex4-2}

\usepackage[utf8]{inputenc}
\usepackage{graphicx}
\usepackage{hyperref}
\usepackage{xcolor,comment}
\usepackage{natbib}
\usepackage[T1]{fontenc}

\usepackage{amsmath,amsthm,amssymb}
\usepackage{mathtools}
\usepackage{braket}
\usepackage{bbm,bm}
\usepackage[bb=boondox]{mathalfa}

\newcommand{\ie}{{i.e.},\ }

\newcommand{\Tr}{\operatorname{Tr}}

\def\ent{S}

\def\spin{J} 
\def\sd{j} 
\def\ms{\mathfrak{j}} 
\def\vol{L} 
\def\cnn{\lambda} 
\def\cnnn{\lambda'} 

\begin{document}
\title{Average pure-state entanglement entropy in spin systems with SU(2) symmetry}

\author{Rohit Patil}
\affiliation{Department of Physics, The Pennsylvania State University, University Park, Pennsylvania 16802, USA}

\author{Lucas Hackl}
\affiliation{School of Mathematics and Statistics, The University of Melbourne, Parkville, VIC 3010, Australia}
\affiliation{School of Physics, The University of Melbourne, Parkville, VIC 3010, Australia}

\author{George R. Fagan}
\affiliation{Department of Physics, The Pennsylvania State University, University Park, Pennsylvania 16802, USA}

\author{Marcos Rigol}
\affiliation{Department of Physics, The Pennsylvania State University, University Park, Pennsylvania 16802, USA}

\begin{abstract}
We study the effect that the SU(2) symmetry, and the rich Hilbert space structure that it generates in lattice spin systems, has on the average entanglement entropy of highly excited eigenstates of local Hamiltonians and of random pure states. Focusing on the zero total magnetization sector ($\spin_z=0$) for different fixed total spin $\spin$, we argue that the average entanglement entropy of highly excited eigenstates of quantum-chaotic Hamiltonians and of random pure states has a leading volume-law term whose coefficient $s_A$ depends on the spin density $\sd=\spin/(\ms\vol)$, with $s_A(\sd \rightarrow 0)=\ln (2\,\ms+1)$ and $s_A(\sd \rightarrow 1)=0$, where $\ms$ is the microscopic spin. We provide numerical evidence that $s_A$ is smaller in highly excited eigenstates of integrable interacting Hamiltonians, which lends support to the expectation that the average eigenstate entanglement entropy can be used as a diagnostic of quantum chaos and integrability for Hamiltonians with non-Abelian symmetries. In the context of Hamiltonian eigenstates we consider spins $\ms=\frac12$ and $1$, while for our calculations based on random pure states we focus on the spin $\ms=\frac12$ case.
\end{abstract}

\maketitle

\section{Introduction}

Entanglement is a foundational concept in quantum mechanics. It provides crucial insights on phenomena that occur across fields in physics, including black-hole evaporation~\cite{page1993information}, quantum phase transitions~\cite{eisert_colloquium_2010}, and quantum dynamics~\cite{kim_huse_13, alba_calabrese_17}. A commonly studied measure of entanglement in pure states is the bipartite entanglement entropy. In strongly interacting quantum many-body systems, the behavior of the bipartite entanglement entropy of highly excited energy eigenstates has become a topic of much current interest (see Ref.~\cite{bianchi2022volume} for a review). In the absence of strong disorder, independent of the integrable or quantum-chaotic nature of the interacting model, the average entanglement entropy of such states generally scales with the volume of the subsystem of interest (when smaller than one-half of the volume of the system)~\cite{bianchi2022volume}. This is to be contrasted to the ``area-law'' scaling of the entanglement entropy of ground states~\cite{eisert_colloquium_2010}. Recently, it was conjectured that the coefficient of the volume in the average entanglement entropy of highly excited energy eigenstates can serve as a diagnostic of quantum-chaos and integrability~\cite{leblond_entanglement_2019}. The coefficient is expected to be maximal in quantum-chaotic systems versus sub-maximal and dependent on the ratio of the subsystem to the system volume in integrable systems.

An analytic understanding of the numerically observed behavior of the average entanglement entropy of highly excited eigenstates of many-body Hamiltonians has been gained using different classes of random states. The (Haar-measure) average entanglement entropy of random states~\cite{page_average_1993} describes the observed leading-order behavior of the average entanglement entropy of quantum-chaotic Hamiltonian eigenstates~\cite{bianchi2022volume}, i.e., like many other properties of such highly excited eigenstates, the leading behavior of their entanglement entropy is described by random matrix theory~\cite{dalessio_quantum_2016}. Differences between the random-state predictions and the numerical results for local Hamiltonian eigenstates have been observed at the level of the subleading $O(1)$ correction~\cite{Haque_Khaymovich_2022, huang_21, kliczkowski2023average, nieva2023}. The volume-law term in the (Haar-measure) average entanglement entropy of random Gaussian states~\cite{lydzba_rigol_20, lydzba_rigol_21, bianchi_hackl_21}, on the other hand, resembles the one in the average entanglement entropy of highly excited eigenstates of integrable interacting Hamiltonians~\cite{bianchi2022volume}. A similar behavior of the leading volume-law term was observed, and rigorous bounds were calculated, for many-body Hamiltonian eigenstates of translationally invariant quadratic models~\cite{vidmar_entanglement_quadratic_2017, vidmar2018volume, hackl_vidmar_19}.  

Another question that has been explored is the role of Abelian symmetries in the behavior of the average entanglement entropy~\cite{bianchi2022volume, bianchi2019typical}. Specifically, the presence of U(1) symmetry in spin-$\frac12$ models (particle-number conservation in spinless-fermion models) was shown to introduce a first subleading correction to the average entanglement entropy that depends on the square root of the volume~\cite{vidmar_entanglement_2017}. Remarkably, the same first subleading correction was found in random pure states with fixed total magnetization or particle number~\cite{vidmar_entanglement_2017, bianchi2022volume}. Energy conservation was later argued to have a similar effect in Hamiltonian eigenstates~\cite{murthy_19}. 

In this work we explore the effect that the non-Abelian SU(2) symmetry has on the average entanglement entropy of highly excited eigenstates of local Hamiltonians and of random pure states in lattice systems. Non-Abelian symmetries are present in models studied across fields in physics~\cite{kogut_79}. Recently, they have attracted significant attention in the context of quantum-information thermodynamics~\cite{guryanova_16, halpern_16, popescu_20, manzano_22, halpern_23}, and they have been identified as a route to generating quantum many-body scars~\cite{Moudgalya_2022, chandran_23}. Recent studies have also explored the effect that such symmetries have on the eigenstate thermalization hypothesis~\cite{murthy_non-Abelian_2023, noh_eigenstate_2023}.

We use numerical simulations to study the average entanglement entropy of highly excited eigenstates of quantum-chaotic and integrable interacting Hamiltonians with SU(2) symmetry. Our goal is to find how the average entanglement entropy of energy eigenstates with different total angular momentum scales with the volume of the subsystem of interest, and whether quantum-chaotic and integrable Hamiltonian eigenstates exhibit different behaviors (as they do in models without symmetries or with Abelian symmetries). A second goal of our work is to carry out analytic calculations of the average entanglement entropy of random pure states that are eigenstates of the SU(2) related conserved quantities to determine whether such averages describe the behavior observed numerically for the Hamiltonian eigenstates of the quantum-chaotic models.

We focus on pure states with zero total magnetization ($\spin_z=0$), and compute the average entanglement entropy for different fixed values of the total spin $\spin$ (and number of lattice sites $\vol$). We argue that the average entanglement entropy of highly excited eigenstates of quantum-chaotic Hamiltonians and of random pure states has a coefficient $s_A$ of the volume that depends on the spin density $\sd=\spin/(\ms \vol)$, with $s_A\rightarrow \ln d$ as $\sd \rightarrow 0$ and $s_A\rightarrow 0$ as $\sd \rightarrow 1$, where $\ms$ is the microscopic spin (notice the difference with the italic $j$ used for the spin density) and $d=2\,\ms+1$ is the size of the Hilbert space of a lattice site. For highly excited eigenstates of integrable interacting Hamiltonians, on the other hand, we provide numerical evidence that $s_A$ is smaller than for quantum-chaotic Hamiltonians. We report numerical results for eigenstates of spin $\ms=\frac12$ and $1$ Hamiltonians, while for our analytic and numerical calculations involving random pure states we focus on the spin $\ms=\frac12$ case.

The presentation is organized as follows. In Sec.~\ref{sec:u1}, we introduce the setup for our study of the entanglement entropy and review previous results in the absence and presence of U(1) symmetry. The dimensions of the sectors of the Hilbert space in the presence of SU(2) symmetry are discussed in Sec.~\ref{sec:hilbspacdim}. We introduce the Hamiltonians considered and report results for the average entanglement entropy of their highly excited eigenstates in Sec.~\ref{sec:hamiltonias}. Section~\ref{sec:randomstates} is devoted to the study of the average entanglement entropy of random pure states. A summary and discussion of our results is provided in Sec.~\ref{sec:summary}.

\section{Entanglement entropy and\\ U(1) symmetry}\label{sec:u1}

We study the bipartite entanglement entropy of pure states $\ket{\psi}\in \mathcal{H}^\ms$ of $\ms$ spins in a lattice with $\vol$ sites, where 
\begin{equation}\label{eq:totalHS}
 \mathcal{H}^\ms=(\ms)^{\otimes \vol},
\end{equation}
for bipartitions 
\begin{equation}
\mathcal{H}^\ms=\mathcal{H}^\ms_A\otimes\mathcal{H}^\ms_B=(\ms)^{\otimes \vol_A}\otimes (\ms)^{\otimes \vol_B}, 
\end{equation}
involving $\vol_A$ ($\vol_B$) contiguous $\ms$ spins in the subsystem of interest $A$ (the complement $B$), with $\vol=\vol_A+\vol_B$. The entanglement entropy of subsystem $A$ is
\begin{equation}
   \ent_A(\ket{\psi})=-\mathrm{\Tr}(\hat \rho_A \ln \hat \rho_A) ,
\end{equation}
where a mixed
\begin{equation}
\hat \rho_A=\mathrm{\Tr}_B(\ket{\psi}\bra{\psi})
\end{equation}
is obtained after tracing out the complement $B$. 

The (Haar-measure) average entanglement entropy of random pure states in such systems is known to be nearly maximal. It has, for $\vol_A\leq \vol/2$, the form~\cite{page_average_1993}
\begin{equation}\label{eq:page_lead}
\langle S_A\rangle=\vol_A \ln d  - \frac{1}{2} \delta_{f,\frac{1}{2}} + o(1),
\end{equation}
where $d=2\,\ms+1$, $f=\vol_A/\vol$ is what we call the ``subsystem fraction,'' and we use $o(1)$ to refer to terms that vanish in the thermodynamic limit (Landau's little $o$ notation). 
The result for $\vol_A>\vol/2$ ($f>\frac12$) follows after replacing $\vol_A\rightarrow \vol-\vol_A$ in Eq.~\eqref{eq:page_lead}. Note that, in Eq.~\eqref{eq:page_lead}, the leading volume-law term is maximal. $\langle S_A\rangle$ in Eq.~\eqref{eq:page_lead} is not maximal because of the $O(1)$ correction ($-\frac12$) that appears at the subsystem fraction $f=\frac12$. 

To understand how symmetries present in Hamiltonians of interest change the average entanglement entropy of highly excited energy eigenstates, one can carry out (Haar-measure) averages of the entanglement entropy of random pure states that are eigenstates of the conserved quantities associated to those symmetries. Before discussing the case of SU(2) symmetry, our interest here, we summarize previous results for the U(1) case. The conserved quantity associated to the U(1) symmetry is the total magnetization $\hat \spin_z$, which is also conserved in the presence of the (higher) SU(2) symmetry. 

For spin $\ms=\frac12$ systems with U(1) symmetry, as mentioned before, in Refs.~\cite{vidmar_entanglement_2017, bianchi2022volume} it was shown that fixing the total magnetization when carrying out the averages introduces a subleading correction that scales with the square root of $\vol_A$. These results are also of relevance to spinless fermion systems with particle number conservation, in which the total particle number $N$ plays the role that the total magnetization $\spin_z$ plays for spin $\ms=\frac12$ systems, $N=\spin_z + \vol/2$. When (more conveniently) written in terms of the fermion filling $n=N/\vol$, which is equivalent to the total magnetization per site $\sd_z=\spin_z/\vol$, $n=\sd_z+\frac12$, the (Haar-measure) average entanglement entropy of random pure states with fixed $N$ has the form~\cite{vidmar_entanglement_2017, bianchi2022volume}:
\begin{align}\label{eq:n_lead}
	\langle S_A\rangle_{n}=&-[n\ln n+(1-n)\ln(1-n)]\, \vol_A\nonumber\\
	&-\sqrt{\frac{n(1-n)}{2\pi}}\left|\ln\left(\frac{1-n}{n}\right)\right|\delta_{f,\frac{1}{2}}\sqrt{\vol}\nonumber\\
	&+\frac{f+\ln(1-f)}{2}-\frac{1}{2}\delta_{f,\frac{1}{2}}\delta_{n,\frac{1}{2}}+o(1),
\end{align}
for $\vol_A\leq \vol/2$. The result for $\vol_A>\vol/2$ ($f>\frac12$) follows from Eq.~\eqref{eq:n_lead} after replacing $\vol_A\rightarrow \vol-\vol_A$. Three points to emphasize about $\langle S_A\rangle_{n}$ in Eq.~\eqref{eq:n_lead} are as follows: (i) The coefficient of $\vol_A$ in the first term depends on $n$~\cite{vidmar_entanglement_2017, grover18} and agrees with the one in Eq.~\eqref{eq:page_lead} at $n=\frac12$; (ii) the coefficient of $\sqrt{\vol_A}$ in the second term vanishes at $n=\frac12$, i.e., it is only at half-filling that there is no square-root-of-the-volume correction; and (iii) the $O(1)$ correction has one term that depends only on $f$~\cite{vidmar_entanglement_2017} and a $-\frac12$ that appears only at $n=\frac12$ and $f=\frac12$~\cite{bianchi2022volume}.

Equation~\eqref{eq:n_lead} is a result of the fact that the Hilbert space $\mathcal{H}^{(N)}$ of the system at a fixed eigenvalue $N$ of $\hat{N}$ is a direct sum of tensor products
\begin{align}\label{eq:fixedN}
	\mathcal{H}^{(N)}=\bigoplus^{N}_{N_A=0}\left(\mathcal{H}_A^{(N_A)}\otimes\mathcal{H}_B^{(N-N_A)}\right)\,,
\end{align}
with $N_A$ being the eigenvalues of $\hat{N}$ in subsystem $A$. While the full derivation of Eq.~\eqref{eq:n_lead} is lengthy (see Ref.~\cite{bianchi2022volume}), the leading volume-law term can be advanced as follows. The Hilbert space of a system with $\vol_A$ sites and $N_A$ spinless fermions is 
\begin{equation}\label{eq:ferhilbspace}
    D_{N_A}\equiv \dim \mathcal{H}^{(N_A)}\,=\,\binom{\vol_A}{N_A} .
\end{equation}
Using Stirling's approximation for the particular case in which $N_A/\vol_A=N/\vol=n$, one can write
\begin{equation}
    D_{N_A} \simeq \frac{1}{\sqrt{2\pi \vol_A}\sqrt{n(1-n)}} \exp\left(-\ln\left[n^n (1-n)^{1-n}\right]\vol_A\right).\label{eq:DNA-asymp}
\end{equation}
The leading term in Eq.~\eqref{eq:n_lead}, is the same as the leading term in $\ln D_{N_A}$ for $N_A/\vol_A=N/\vol=n$, i.e., it is the same as the leading volume-law term of the logarithm of the Hilbert space dimension of subsystem $A$ at the same average site occupation as the entire system. This is equivalent to taking the reduced density matrix $\hat\rho_A$ of subsystem $A$ to be that of a maximally mixed state of $N_A=n\vol_A$ fermions in $\vol_A$ sites. For $\vol_A > \vol/2$, the relevant maximally mixed state is the one in the complement of $A$, with $\vol_B=\vol-\vol_A$ sites and $N_B=n\vol_B$ fermions. 

\section{Entanglement entropy and\\ SU(2) symmetry}\label{sec:hilbspacdim}

To account for the presence of SU(2) symmetry, one can carry out (Haar-measure) averages of the entanglement entropy of random states that are simultaneous eigenstates of $\hat{\vec \spin}^{\,2}$ and $\hat \spin_z$. In this section we discuss the dimensions of such sectors of the Hilbert space and what those dimensions advance about the leading volume-law term of the average entanglement entropy of the corresponding random states.

The representation theory of $\mathrm{SU}(2)$ allows us to rewrite the $L$th tensor product of the spin-$\ms$ representation in Eq.~\eqref{eq:totalHS} as a direct sum 
\begin{equation}
    \mathcal{H}^\ms=\bigoplus^{\ms\vol}_{\spin=\spin_{\min}} \underbrace{\spin\oplus\dots\oplus \spin}_{n^\ms_\spin\text{ times}}\equiv \bigoplus^{\ms\vol}_{\spin=\spin_{\min}} \mathcal{H}^\ms_\spin \,,\label{eq:H-decomposition}
\end{equation}
where the sum runs over integer (half-integer) spins $\spin$ starting at $\spin_{\min}=0$ ($\spin_{\min}=\frac12$ for $L>1$) for even (odd) $2\ms\vol$, and $n^{\ms}_{\spin}$ is the multiplicity of a spin $\spin$.

For large $\vol$, we can express $\spin$ in terms of the spin density
\begin{align}
    \sd=\frac{\spin}{\ms \vol}\,,
\end{align}
which allows us to write the asymptotic form of the multiplicities in the form
\begin{align}
    n^{\ms}_{\spin}\simeq \frac{\alpha^\ms(j)}{\sqrt{\vol}}\exp\left[\beta^\ms(j)\vol\right]\,,\label{nJ-asymp}
\end{align}
where the functions $\alpha^\ms(j)$ and $\beta^\ms(j)$ can be computed using the group theory method of Weyl characters~\cite{hall2013lie} (see Appendix~\ref{sec:AppHSD}). The key result is that $\beta^\ms(j)=\psi_\ms(z_0(j))$ can be found as a saddle point, where $\psi_\ms(z)$ is given in Eq.~\eqref{eq:psi-def} and $z_0\geq 0$ is the unique non-negative real solution of the saddle point equation $\psi'_\ms(z_0)=0$.

The dimension of the Hilbert space sector with fixed $J$ and the one with fixed $(J,J_z)$ are given by
\begin{align}
    D^{\ms}_{\spin}&=\dim\mathcal{H}^\ms_{\spin}=(2\spin+1)\,n^{\ms}_{\spin}\,,\\
    D^{\ms}_{\spin,\spin_z}&=\dim\mathcal{H}^\ms_{\spin,\spin_z}=\begin{cases}
        n^{\ms}_{\spin} & |J_z|\leq J\\
        0 & \text{else}
    \end{cases}\,.
\end{align}
We therefore see that both will have the same exponential scaling from Eq.~\eqref{nJ-asymp} encoded in $\beta^\ms(j)$. 

Drawing the analogy to $D_{N_A}$ in Eq.~\eqref{eq:DNA-asymp} and $\braket{S_A}$ in Eq.~\eqref{eq:n_lead}, we thus expect that the leading order behavior of the average entanglement entropy of random pure states at fixed $\spin$ will be given by
\begin{align}\label{eq:leadaverbeta}
    \braket{S_A}^\ms_{\spin}=\beta^\ms(j)\vol_A+o(\vol_A)\,,
\end{align}
regardless of whether we fix $\spin_z$ or not. 

We focus next on the microscopic spin values $\ms=\frac12$ and $\ms=1$ restricted to the zero total magnetization sector $\spin_z=0$, for which we carry out numerical calculations of the average entanglement entropy of highly excited eigenstates of quantum-chaotic and integrable interacting Hamiltonians in the next section. 

\begin{figure}[!t]
    \includegraphics[width=0.98\columnwidth]{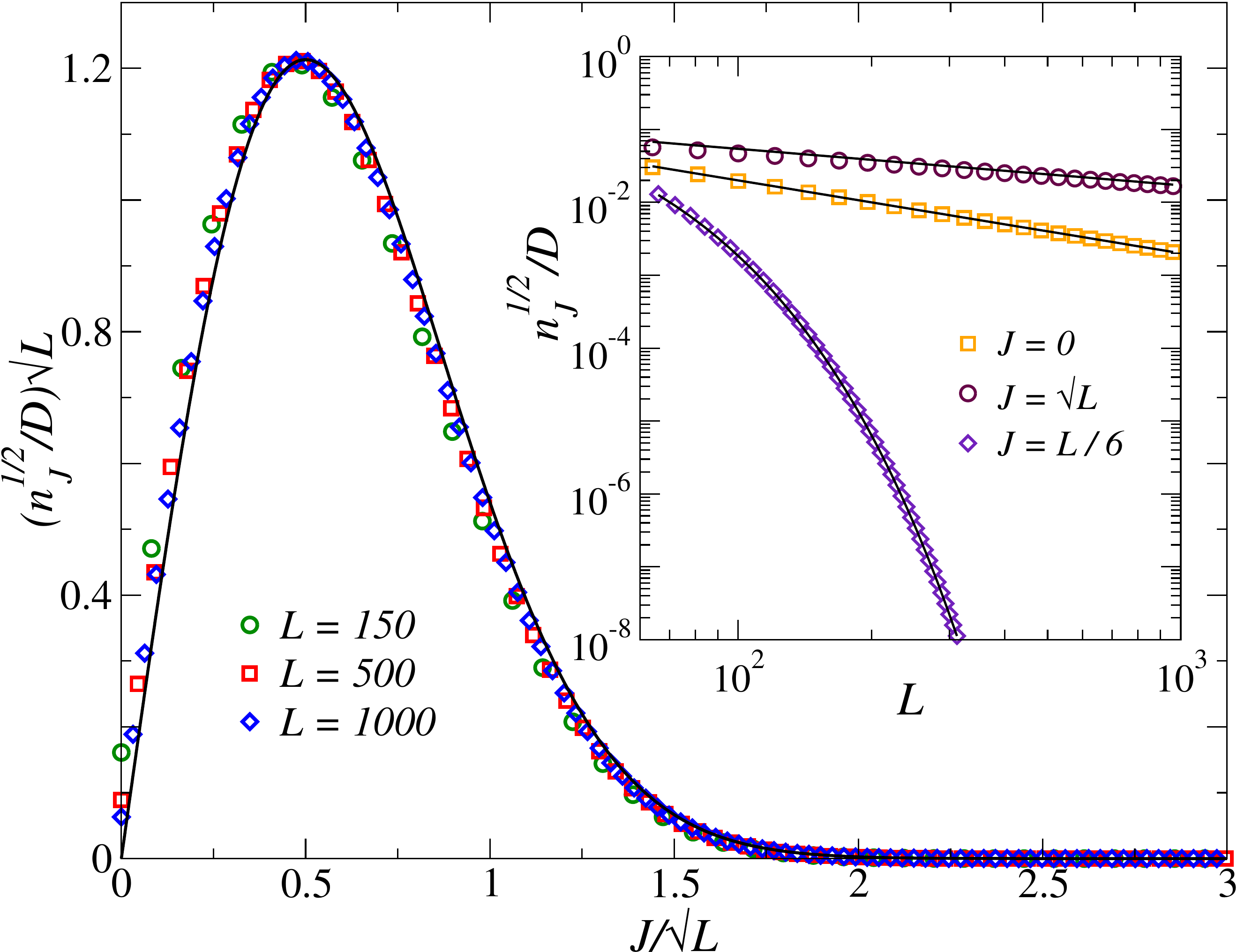}
    \caption{{\it Hilbert space fraction $n^{1/2}_\spin/D$}. (Main panel) Results predicted by Eq.~\eqref{eq:HilbSpaceDimension} vs $\spin$, rescaled using $\sqrt{\vol}$, for $L=150$, 500, and 1000. The solid line shows $\frac{n^{1/2}_\spin}{D}\sqrt{\vol}=\frac{4\spin}{\sqrt{\vol}}e^{-2\spin^2/\vol}$, obtained from Eq.~\eqref{eq:AsymptoticHilbFraction} as the leading order for $\spin=O(\sqrt{\vol})$. (Inset) Results predicted by Eq.~\eqref{eq:HilbSpaceDimension} vs $L$ for $\spin=0,\,\sqrt{\vol},$ and $\vol/6$. The solid lines show the leading order predicted by Eq.~\eqref{eq:AsymptoticHilbFraction} in each regime, where $n^{1/2}_\spin/D=a\vol^{-1},\,b\vol^{-1/2}$, and $ce^{-d\vol}$, with $a,\,b,\,c,\,d=2,\,0.54,\,0.53,\,0.057$, respectively.}
    \label{fig:HilbertSpaceFraction}
\end{figure}

\textbf{Spin $\ms=\frac12$.} The multiplicity $n^{1/2}_J$ can be calculated exactly for finite systems using the closed form expression of an integral in Appendix~\ref{sec:AppHSD}, or the combinatorics approach explained in Appendix~\ref{sec:S2},
\begin{equation}\label{eq:HilbSpaceDimension}
        n^{1/2}_\spin=\frac{2(1+2\spin)}{2+\vol+2\spin}\binom{\vol}{\vol/2-\spin}\,.
\end{equation}
When written in the asymptotic form in Eq.~\eqref{nJ-asymp}, one finds that $\beta^{1/2}(j)$ has the form
\begin{equation}\label{eq:AsymptoticHilbSpcDim}
 \beta^{1/2}(j)=-\left[ \frac{1+j}{2} \ln \left(\frac{1+j}{2}\right) + \frac{1-j}{2} \ln \left(\frac{1-j}{2}\right) \right].
\end{equation}
$\beta^{1/2}(j)$ will be used in our comparison to the numerical results obtained for the average entanglement entropy of highly excited Hamiltonian eigenstates in Sec.~\ref{sec:hamiltonias}, and to the analytical results obtained for the average entanglement entropy of random pure states in Sec.~\ref{sec:randomstates}. 

In Fig.~\ref{fig:HilbertSpaceFraction}, we show the rescaled fraction of states $n^{1/2}_\spin/D \times \sqrt{\vol}$ with spin $\spin$ in the zero magnetization sector vs the rescaled $\spin/\sqrt{\vol}$ for three values of $\vol$. We refer to the Hilbert space of the full $\spin_z=0$ sector as $D$, which can be obtained using Eq.~\eqref{eq:ferhilbspace} replacing $\vol_A\rightarrow \vol$ and $N_A\rightarrow \vol/2$. As $\vol$ increases, the rescaling used produces a collapse of the curves for different values of $\vol$, with the maximal $n^{1/2}_\spin/D\approx 1.2/\sqrt{\vol}$ for $\spin \approx \sqrt{\vol}/2$. The collapse in Fig.~\ref{fig:HilbertSpaceFraction} makes apparent that as $\vol$ increases the sectors with $\spin=O(\sqrt{\vol})$ account for an increasingly large fraction of the entire Hilbert space. 

The inset in Fig.~\ref{fig:HilbertSpaceFraction} shows the scaling of $n^{1/2}_\spin/D$ with $\vol$ for: $\spin=O(1)$, $n^{1/2}_\spin/D \propto 1/\vol$; $\spin=O(\sqrt{\vol})$, $n^{1/2}_\spin/D \propto 1/\sqrt{\vol}$; and $\spin=O(\vol)$, $n^{1/2}_\spin/D$ decays exponentially with $\vol$. Those scalings can be obtained analytically using that, for large values of $L$, $n^{1/2}_\spin/D$ can be written as
\begin{eqnarray}\label{eq:AsymptoticHilbFraction}
    \frac{n^{1/2}_\spin}{D} &\simeq&  \frac{2}{\sqrt{1-\sd^2}}\left(\frac{\sd}{1+\sd}+\frac{1-\sd}{(1+\sd)^2\vol}\right)\, \times\,  \\ & &\exp\left(-\left[\frac{1+\sd}{2}\ln(1+\sd)+\frac{1-\sd}{2}\ln(1-\sd)\right]\vol\right)\,, \nonumber 
\end{eqnarray}
where $\sd=\frac{2\spin}{\vol}$ is the spin density ($\sd\in[0,1]$). When $\spin=O(\sqrt{\vol})$, one obtains $\frac{n^{1/2}_\spin}{D}\sqrt{\vol}=\frac{4\spin}{\sqrt{\vol}}e^{-2\spin^2/\vol}$, which describes the results in Fig.~\ref{fig:HilbertSpaceFraction} for large values of $\vol$. One can solve for the location of the maximum in $n^{1/2}_\spin/D$, from $d(n^{1/2}_\spin/D)/d\sd=0$, via the transcendental equation
\begin{align}
    \vol[1+\sd(\sd-1)-\sd(1-\sd^2)\mathrm{arctan}(\sd)\vol]=0\,.
\end{align}
This equation can be solved perturbatively in the limit $\vol\to \infty$. We find $\sd=\frac{1}{\sqrt{\vol}}-\frac{1}{2\vol}+\frac{9}{8\vol^{3/2}}+O(1/\vol^2)$, which is where the maximum was identified in Fig.~\ref{fig:HilbertSpaceFraction}.

\textbf{Spin $\ms=1$.} For the microscopic spin $\ms=1$ case, we focus solely on the asymptotic behavior of dimensions of the Hilbert spaces of interest. As discussed in Appendix~\ref{sec:AppHSD}, for $\ms=1$ one finds that $\beta^1(j)$ [see Eq.~\eqref{nJ-asymp}] takes the form
\begin{align}
    \beta^1(j)=\ln\tfrac{3}{\sqrt{4-3j^2}-1}+j\ln\tfrac{\sqrt{4-3j^2}-j}{2(1+j)}\,.\label{eq:beta-frakj=1}
\end{align}
$\beta^1(j)$ will be used in our comparison to the results obtained for the average entanglement entropy of highly excited Hamiltonian eigenstates in the next section.

\section{SU(2)-symmetric Hamiltonians}\label{sec:hamiltonias}

We study the spin $\ms =\frac12$ extended Heisenberg model with nearest and next-nearest (with strength $\cnn$) neighbor interactions in chains with $L$ sites   
\begin{equation}\label{eq:Hamiltonian Spin 1/2}
    H=-\sum_{i=1}^\vol \hat{\vec{\spin}}_i \cdot \hat{\vec{\spin}}_{i+1} - \cnn \sum_{i=1}^\vol  \hat{\vec{\spin}}_i \cdot \hat{\vec{\spin}}_{i+2}\,,  
\end{equation}
where $\hat{\vec{\spin}}_i=(\hat{\spin}^x_i,\hat{\spin}^y_i,\hat{\spin}^z_i)$ is the spin-$\frac12$ operator at site $i$, and we use periodic boundary conditions. This model is integrable when $\cnn=0$, and quantum chaotic (nonintegrable) when $\cnn\neq0$ (we set $\cnn=3$ in the latter regime, see Appendix~\ref{sec:S1}). 

We also study the spin $\ms =1$ extended Heisenberg model with nearest-neighbor interactions in a chain of $L$ sites with the Hamiltonian
\begin{equation}\label{eq:Hamiltonian Spin 1}
    H'=-\sum_{i=1}^\vol \hat{\vec{\spin}}_i' \cdot \hat{\vec{\spin}}_{i+1}' + \cnnn \sum_{i=1}^\vol  (\hat{\vec{\spin}}_i' \cdot \hat{\vec{\spin}}_{i+1}')^2\,,  
\end{equation}
where $\hat{\vec{\spin}}_i'=(\hat{\spin}^{'x}_i,\hat{\spin}^{'y}_i,\hat{\spin}^{'z}_i)$ is the spin-$1$ operator at site $i$, also with periodic boundary conditions. As opposed to its $\ms=\frac12$ counterpart with $\lambda=0$, the $\ms =1$ Heisenberg model in Eq.~\eqref{eq:Hamiltonian Spin 1} with $\cnnn=0$ (i.e., with only the first term in the sum) is quantum chaotic. The second term with $\cnnn=1$ makes the model integrable~\cite{zamolodchikov1980model,BABUJIAN1982479}.

The Hamiltonians in Eq.~\eqref{eq:Hamiltonian Spin 1/2} and~\eqref{eq:Hamiltonian Spin 1} are translationally invariant so the total quasimomentum is conserved. We compute the average entanglement entropy of sectors with different fixed spin $\spin$ ($\spin_z=0$) using the central $20\%$ of the energy eigenstates in the total quasi-momentum subsectors $k_n=2\pi n/L$ with $n=1,\,2,\,\ldots,L/2-1$. The results reported are the averages $\bar \ent_A$ over all those ``complex'' sectors~\cite{leblond_ethsymm_2020, kliczkowski2023average}. 

\begin{figure}[!t]
    \includegraphics[width=0.98\columnwidth]{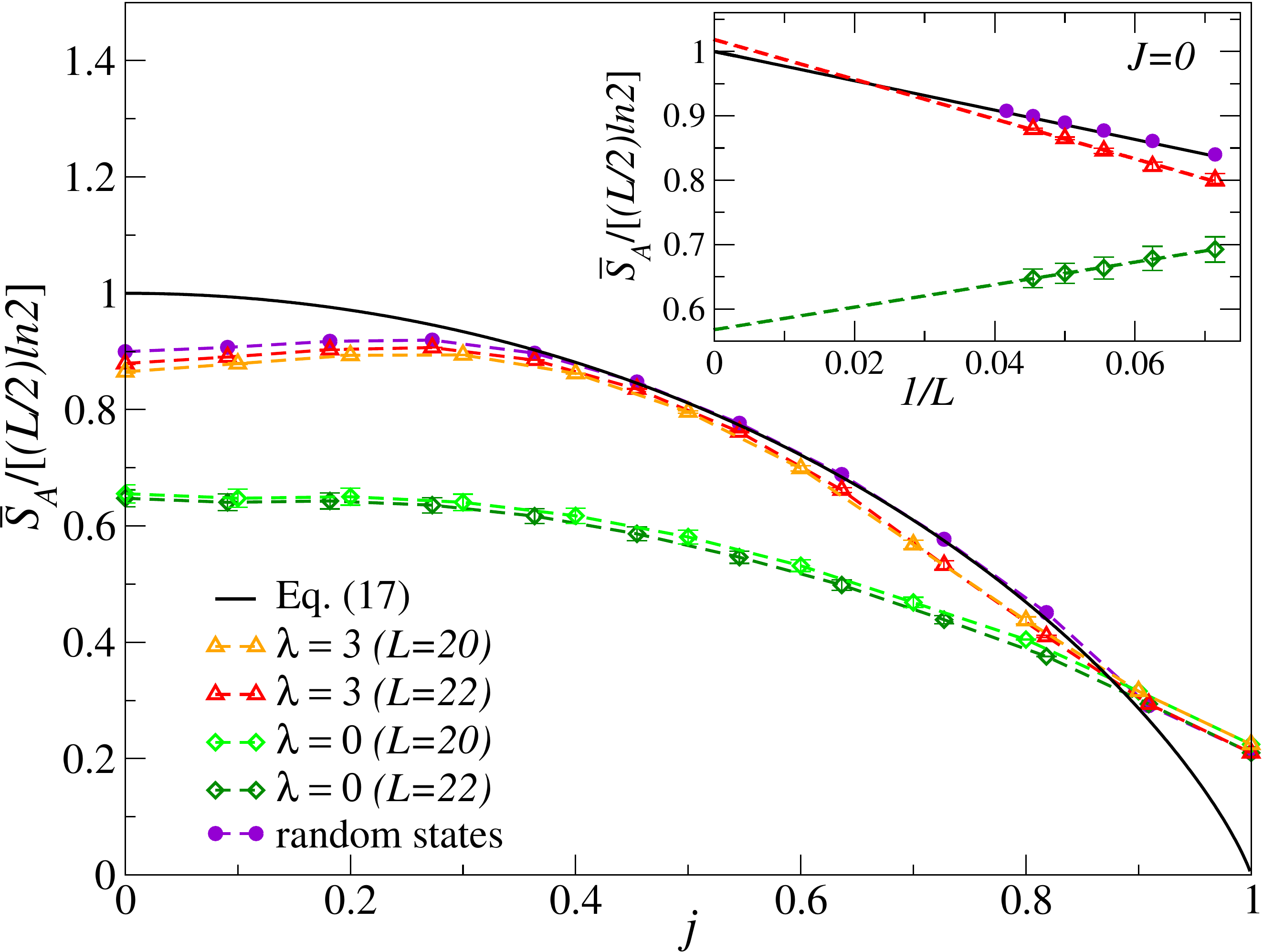}
    \caption{{\it Average entanglement entropy $\bar \ent_A$ at $f=\frac12$ for the $\ms=\frac12$ Hamiltonian in Eq.~\eqref{eq:Hamiltonian Spin 1/2}.} $\bar \ent_A$ vs $\sd=2\spin/\vol$ for quantum-chaotic ($\cnn=3$) and integrable ($\cnn=0$) Hamiltonian eigenstates in systems with $\vol=20$ and $22$. We also show results for the average entanglement entropy of random pure states for $L=22$ [see Eq.~\eqref{eq:randavenum}]. The continuous line shows the prediction for $\braket{S_A}^{1/2}_{\spin}$ from Eq.~\eqref{eq:leadaverbeta} with $\beta^{1/2}(j)$ from Eq.~\eqref{eq:AsymptoticHilbSpcDim}. (inset) $\bar \ent_A$ vs $1/\vol$ at $\spin=0$, for quantum-chaotic and integrable Hamiltonian eigenstates, as well as for random states. The solid line shows the prediction for $\braket{S_A}_{\spin=0,\spin_z=0}$ from Eq.~\eqref{eq:S=0_Asymptotic}. The error bars are the standard deviation of the averages, and the dashed lines are linear ($a+b/\vol$) fits to the Hamiltonian data shown.}
    \label{fig:Spin 1/2 per site}
\end{figure}

\begin{figure}[!t]
    \includegraphics[width=0.98\columnwidth]{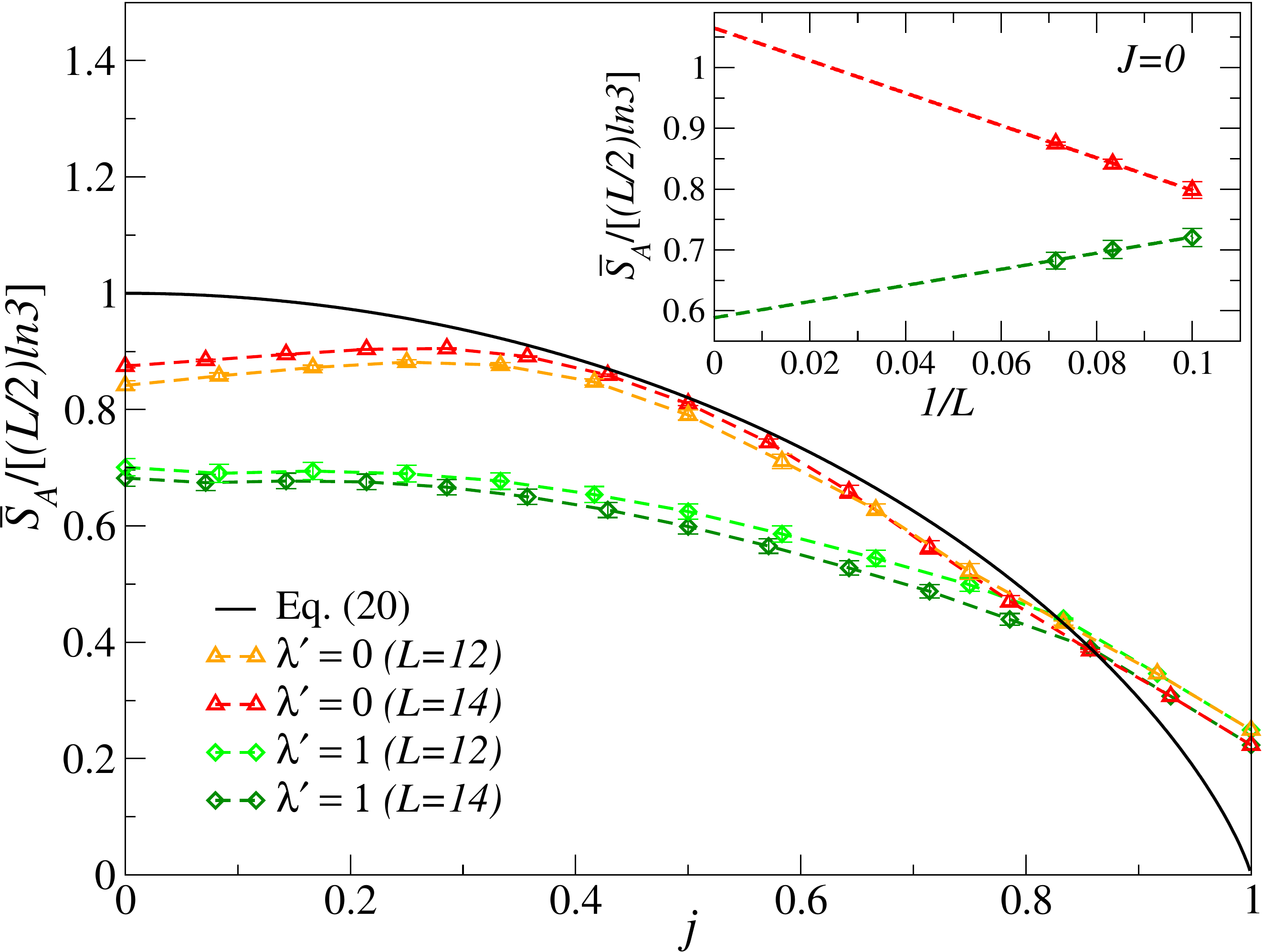}
    \caption{{\it Average entanglement entropy $\bar \ent_A$ at $f=\frac12$ for the $\ms=1$ Hamiltonian in Eq.~\eqref{eq:Hamiltonian Spin 1}.} $\bar \ent_A$ vs $\sd=\spin/\vol$ for quantum-chaotic ($\cnnn=0$) and integrable ($\cnnn=1$) Hamiltonian eigenstates, in systems with $\vol=12$ and $14$. The continuous line shows the prediction for $\braket{S_A}^1_{\spin}$ from Eq.~\eqref{eq:leadaverbeta} with $\beta^{1}(j)$ from Eq.~\eqref{eq:beta-frakj=1}. (inset) $\bar \ent_A$ vs $1/\vol$ at $\spin=0$, for quantum-chaotic and integrable Hamiltonian eigenstates. The error bars are the standard deviation of the averages, and the dashed lines are linear ($a+b/\vol$) fits to the points shown.}
    \label{fig:Spin 1 per site}
\end{figure}

In Fig.~\ref{fig:Spin 1/2 per site} (Fig.~\ref{fig:Spin 1 per site}), we plot $\bar \ent_A$ vs $j$ at subsystem fraction $f=\frac12$ for the eigenstates of the extended Heisenberg model in Eq.~\eqref{eq:Hamiltonian Spin 1/2} [Eq.~\eqref{eq:Hamiltonian Spin 1}] with $\ms =\frac12$ ($\ms=1$). We show results for two system sizes both for the quantum-chaotic and integrable points considered. We plot as a continuous line in Fig.~\ref{fig:Spin 1/2 per site} [Fig.~\ref{fig:Spin 1 per site}] the prediction for $\braket{S_A}^\ms_{\spin}$ from Eq.~\eqref{eq:leadaverbeta} with $\beta^{1/2}(j)$ [$\beta^{1}(j)$] from Eq.~\eqref{eq:AsymptoticHilbSpcDim} [Eq.~\eqref{eq:beta-frakj=1}]. The numerical results for $\bar \ent_A$ for the quantum-chaotic points are distinct from their integrable counterparts away from $\sd=1$ (At maximal total spin the Hilbert space consists of a single state.). One can also see in Figs.~\ref{fig:Spin 1/2 per site} and~\ref{fig:Spin 1 per site} that, with increasing $\vol$, $\bar \ent_A$ for quantum-chaotic energy eigenstates approaches the $\braket{S_A}^\ms_{\spin}$ predictions, while $\bar \ent_A$ for integrable energy eigenstates departs from the $\braket{S_A}^\ms_{\spin}$ predictions away from $\sd=1$. 

The aforementioned scaling behaviors with increasing system size are better seen in the insets in Figs.~\ref{fig:Spin 1/2 per site} and~\ref{fig:Spin 1 per site}, in which we show finite-size scaling analyses for $\bar \ent_A$ at $f=\frac12$ and $\spin=0$. Both for $\ms=\frac12$ and 1, we find evidence that $\bar \ent_A$ has a leading volume-law term no matter whether the model is integrable or quantum chaotic. For the quantum-chaotic energy eigenstates we find that the coefficient of the volume law $s_A=\lim_{L\rightarrow \infty}\bar \ent_A/(\vol/2)$ is consistent with the maximal value of $\ln d$ (with $d=2\,\ms+1$) as predicted by Eqs.~\eqref{eq:AsymptoticHilbSpcDim} and~\eqref{eq:beta-frakj=1} for $\sd=0$. For the integrable energy eigenstates, on the other hand, $s_A$ appears to be only slightly larger than one-half of the maximal value, as found in Ref.~\cite{leblond_entanglement_2019} for the spin-$\frac12$ XXZ model, which has only U(1) symmetry.  

Having seen that the results for the leading volume-law term of the average entanglement entropy of highly excited eigenstates of SU(2) symmetric quantum-chaotic Hamiltonians with $\ms=\frac12$ and 1 are consistent with the prediction from a maximally mixed state in the relevant sector of the Hilbert space in subsystem A, in what follows we use analytical and further numerical calculations to address two questions. The first one is whether the direct calculation of the average entanglement entropy of random pure states produces the same leading volume-law term as the maximally mixed state advances. Our intuition on this matter was built based on the results reviewed for the case of U(1) symmetry in Sec.~\ref{sec:u1}, so we need to verify that this intuition also applies for the SU(2) symmetry. The second question we address is the nature of the subleading corrections depending on the value of $J$. Our numerical calculations for Hamiltonian eigenstates are restricted to system sizes that are too small to gain an understanding of how the subleading corrections change depending on the value of $J$, so an analytical treatment is needed to address this question. If the nonvanishing (in the thermodynamic limit) subleading corrections for the average entanglement entropy of highly excited quantum-chaotic Hamiltonian eigenstates have the same form as those for random pure states, as is the case for the U(1) symmetry, then our analytical results for random pure states will provide insights into what is to be expected for the subleading corrections in Hamiltonian eigenstates.

For the analytic calculations in the rest of this work we focus on the spin $\ms=\frac12$ case so, to lighten the notation, we drop $\ms$ from all the expressions that follow. A first indication that the leading behavior of the average entanglement entropy of highly excited quantum-chaotic energy eigenstates behaves similarly to that of the average entanglement entropy of random pure states is provided by the closeness of both averages in Fig.~\ref{fig:Spin 1/2 per site} for $L=22$. The purple circles in Fig.~\ref{fig:Spin 1/2 per site} show our numerical results for the average over random states with spin $\spin$ ($\spin_z=0$). The random pure states are taken to have the form 
\begin{equation}\label{eq:randavenum}
  \ket{\psi}_\text{rand}=\sum_i\ C_i \ket{\spin,0}_i,  
\end{equation}
where $\{\ket{\spin,0}_i\}_{i=1}^{n_\spin}$ is a basis generated by the eigenstates of $\hat{\vec \spin}^{\,2}$ and $\hat\spin_z$ with eigenvalues $\spin$ and $\spin_z=0$, respectively. The random coefficients $C_i \in \mathbb{R}$ are drawn from a normal distribution, and they are normalized to satisfy $\sum_i C_i^2=1$. The scaling of the average over random pure states is shown in the inset in Fig.~\ref{fig:Spin 1/2 per site} for $\spin=0$, which one can see is qualitatively similar to that of the quantum-chaotic energy eigenstates, and follows the analytical prediction for $\braket{S_A}_{\spin=0,\spin_z=0}$ from Eq.~\eqref{eq:S=0_Asymptotic}. The latter shows that the average entanglement entropy of random pure states in the $J=0$ sector produces the same leading volume-law term as the corresponding maximally mixed state in subsystem $A$.

Since all our analytical calculations are carried out using random coefficients $C_i \in \mathbb{C}$ (so that the states $\ket{\psi}_\text{rand}$ are Haar random in the respective Hilbert space) whereas all our numerical calculations are carried out using random coefficients $C_i \in \mathbb{R}$ (to reduce the computation time), we stress that the difference between the results for $C_i \in \mathbb{R}$ and $C_i \in \mathbb{C}$ is exponentially small in $\vol$ (see Appendix~\ref{sec:randcoeff}), i.e., real vs complex coefficients results in negligible differences in what follows.

\section{Entanglement entropy for fixed $\spin$ and $\ms=\frac12$}\label{sec:randomstates}

To compute the average over random states with $\ms=\frac12$ analytically, we write $\mathcal{H}=\mathcal{H}_A\otimes\mathcal{H}_B$ as a direct sum [see Eq.~\eqref{eq:H-decomposition}],
\begin{equation}\label{eq:DirectSumRep}
    \underbrace{\bigoplus^{\vol/2}_{\spin=0} \mathcal{H}_\spin}_{\mathcal{H}}=\underbrace{\bigoplus^{\vol_A/2}_{\spin_A=\spin_{\min}} \mathcal{H}_{\spin_A}}_{\mathcal{H}_A}\otimes \underbrace{\bigoplus^{\vol_B/2}_{\spin_B=\spin_{\min}} \mathcal{H}_{\spin_B}}_{\mathcal{H}_B}\,,
\end{equation}
where, $\spin_{\min}=0\ (\frac12)$ if $\vol_A$ is even (odd), and $n^A_{\spin_A}(\vol_A)=\dim\mathcal{H}_{\spin_A}$ [$n^B_{\spin_B}(\vol_B)=\dim\mathcal{H}_{\spin_B}$] can be obtained using Eq.~\eqref{eq:HilbSpaceDimension} with $\vol \rightarrow \vol_A$ and $\spin \rightarrow \spin_A$ [$\vol \rightarrow \vol_B$ and $\spin \rightarrow \spin_B$]. Equation~\eqref{eq:DirectSumRep} can be interpreted as pairing, given by the principle of angular momentum addition $|\spin_A-\spin_B|\leq \spin\leq \spin_A+\spin_B$, spins $\spin_A$ and $\spin_B$ within subsystems $A$ and $B$ to produce total spin $\spin$. The range of values of $\spin_A$ and $\spin_B$ (depending on $\vol_A$ and $\vol_B$) that can be paired are
\begin{eqnarray}
    \max[\spin_{\min},\spin-\tfrac{\vol_B}{2}]\leq &\spin_A& \leq\min[\tfrac{\vol_A}{2},\spin+\tfrac{\vol_B}{2}]\,,\label{eq:rangesJA}\\
    \max[\spin_{\min},|\spin-\spin_A|]\leq &\spin_B& \leq \min[\tfrac{\vol_B}{2},\spin+\spin_A]\,.\label{eq:rangesJB}
\end{eqnarray}
Next, we study $\bar \ent_A$ separately for $\spin=0$, $\spin=O(1)$, and $\spin=O(\vol)$. As discussed in the context of Fig.~\ref{fig:HilbertSpaceFraction}, for $\spin=O(1)$ one has that $n_\spin/D \propto 1/\vol$ while for $\spin=O(\vol)$, $n_\spin/D$ decays exponentially with $\vol$.
  
\subsection{Spin $\spin=0$}

We consider first $\spin=0$, which is special as $\spin_B=\spin_A$ in Eq.~\eqref{eq:rangesJB}. The $n_0$-dimensional sector $\mathcal{H}_{\spin=0,\spin_z=0}$ can be represented as a direct sum [see Eq.~\eqref{eq:DirectSumRep}]
\begin{equation}
    \mathcal{H}_{\substack{\spin=0,\\\spin_z=0}}=\bigoplus^{\min[\frac{\vol_A}{2},\frac{\vol_B}{2}]}_{\spin_A=\spin_{\min}} \mathcal{H}^0_{\spin_A}\,,\label{eq:HJ0Jz0}
\end{equation}
where $\mathcal{H}^{0}_{\spin_A}\subset \mathcal{H}_{\spin_A}\otimes \mathcal{H}_{\spin_B}$ contains the $n_{\spin_A}^A\times n_{\spin_B}^B$ states that have identical $\spin_A=\spin_B$ and zero total spin. We can explicitly construct the basis for $\mathcal{H}^{0}_{\spin_A}$ as
\begin{equation}
    \ket{\psi_{ab}}=\sum^{\spin_A}_{m=-\spin_A} c_m(\spin_A) \ket{\spin_A,m}_a\otimes\ket{\spin_A,-m}_b\,,
\end{equation}
where $m$ is the $\spin_z$ eigenvalue within subsystem $A$, $a$ ($b$) labels the $n_{\spin_A}^A$ ($n_{\spin_A}^B$) states with spin $\spin_A$ within subsystem $A$ ($B$), and $c_m(\spin_A)$ is the Clebsch-Gordan (CG) coefficient 
\begin{equation}
   \braket{\spin_A,m;\spin_A,-m|\spin=0,\spin_z=0}=\frac{(-1)^{\spin_A-m}}{\sqrt{1+2\spin_A}}. 
\end{equation}

Hence, the Haar-average entanglement entropy $\braket{\ent_A}_{\spin=0,\spin_z=0}$ over random pure states in the spin sector $\mathcal{H}_{\spin=0,\spin_z=0}$ can be computed using the Haar-average entanglement entropy $\braket{\ent_A}^0_{\spin_A}$ over the restricted subspaces $\mathcal{H}^{0}_{\spin_A}$ via~\cite{bianchi2019typical, bianchi2022volume}
\begin{equation}
    \braket{S_A}_{\substack{\spin=0,\\ \spin_z=0}} = \sum_{\spin_A} \frac{d_{\spin_A}}{d} \left[ \braket{{S_A}}^0_{\spin_A} +\Psi(d+1) - \Psi(d_{\spin_A}+1)\right]\, ,\label{eq:tensor-structure}
\end{equation}
where 
\begin{equation}
    \Psi(x)=\frac{\Gamma'(x)}{\Gamma(x)}
\end{equation}
is the digamma function, $d_{\spin_A}=\dim\mathcal{H}^{0}_{\spin_A}$, and $d=\sum_{\spin_A}d_{\spin_A}=\dim\mathcal{H}_{\spin=0,\spin_z=0}$. 

A random state in the subspace $\mathcal{H}_{\spin_A}^0$ can be written as a superposition of base states $\ket{\psi_{ab}}$, 
\begin{equation}
    \ket{\psi^0_{\spin_A}}=\sum^{n^A_{\spin_A}[\vol_A]}_{a=1}\hspace{+0.1cm}\sum^{n^B_{\spin_A}[\vol_B]}_{b=1} W_{ab} \ket{\psi_{ab}}\,,
\end{equation}
where $W_{ab}$ are random numbers drawn from a fixed trace ensemble $\mathrm{Tr}(WW^\dagger)=1$. The corresponding reduced density matrix $\hat{\rho}_A=\mathrm{Tr}_B\ket{\psi^0_{\spin_A}}\bra{\psi^0_{\spin_A}}$ can be written as
\begin{equation}\label{eq:rho_A J=0}
\hat{\rho}_A=\sum_{a,a',m} R_{aa'}^{(m)} \ket{\spin_A,m}_a\bra{\spin_A,m}_{a'}\,.
\end{equation}
It is block-diagonal over spaces of fixed $m$, and the entries in such blocks are $R_{aa'}^{(m)}=|c_m|^2 (WW^\dagger)_{aa'}$. The eigenvalue distribution is thus the product of the fixed distribution of the CG coefficients $|c_m(\spin_A)|^2=1/(1+2\spin_A)$ and the eigenvalue distribution of $WW^\dagger$ from the fixed-trace ensemble, which is the well-known Page result~\cite{page_average_1993} (for subsystems of dimensions $n_{\spin_A}^A$ and $n_{\spin_A}^B$). The entropy of the product of two distributions is the sum of the entropies of the distributions
{\setlength\arraycolsep{0.2pt}
\begin{eqnarray}
    &&\braket{\ent_A}_{\spin_A}^0=\ent_{\mathrm{CG}}(\spin_A) + \ent_{\mathrm{Page}}(n^A_{\spin_A},n^B_{\spin_A})\, , \label{eq:EntCG+Page} \\
    &&\ent_{\mathrm{CG}}(\spin_A)=-\sum_{m}|c_m(\spin_A)|^2\ln|c_m(\spin_A)|^2=\ln(1+2\spin_A)\,,\nonumber\\ 
    &&\ent_{\mathrm{Page}}(d_A,d_B)=\Psi(d_Ad_B+1)-\Psi[\max(d_A,d_B)+1]\nonumber\\
    &&\hspace{2.7cm}-\frac{\min(d_A,d_B)-1}{2\max(d_A,d_B)}\,.\nonumber
\end{eqnarray}
}Plugging Eq.~\eqref{eq:EntCG+Page} in Eq.~\eqref{eq:tensor-structure}, using that $d_{\spin_A}=n_{\spin_A}^An_{\spin_A}^B$ and $d=n_0$, yields an exact expression for $\braket{\ent_A}_{\spin=0,\spin_z=0}$,
\begin{eqnarray}
    \braket{\ent_A}_{\substack{\spin=0,\\ \spin_z=0}}\hspace{-0.1cm}&=&\hspace{-0.5cm}\sum^{\min[\frac{\vol_A}{2},\frac{\vol_B}{2}]}_{{\spin_A}=\spin_{\min}}\frac{n^A_{\spin_A}n^B_{\spin_A}}{n_0}\left[\Psi(n_0+1)-\Psi(n^B_{\spin_A}+1) \right.\nonumber \\&& \hspace{1.5cm}-\frac{n^A_{\spin_A}-1}{2n^B_{\spin_A}}+\left.\ln(1+2\spin_A)\right],
\end{eqnarray}
where we assumed that $\vol_A\leq \vol_B$, without loss of generality due to the $\vol_A\leftrightarrow \vol_B$ symmetry of $\braket{\ent_A}_{\spin=0,\spin_z=0}$.

We then obtain the asymptotic formula in the limit $\vol\to\infty$ for fixed $f=\vol_A/\vol\leq \frac12$ as follows. First, we extract the asymptotic behavior of the density function
\begin{eqnarray}
    \rho({\spin_A})&=&\frac{n_{\spin_A}^An_{\spin_A}^B}{n_0} \\ &=&\sqrt{\frac{8}{\pi}}\frac{(1+2{\spin_A})^2}{\sqrt{f(1-f)\vol}^3}\,\exp\left[-\frac{2{\spin_A}^2}{f(1-f)\vol}\right] + o(1). \nonumber
\end{eqnarray}
We also extract the asymptotic behavior of
\begin{eqnarray} 
 \hspace{-0.3cm} &&\varphi({\spin_A}) = \braket{S_A}_{\spin_A}^0+\Psi(n_0+1) - \Psi(n_{\spin_A}^An_{\spin_A}^B+1) \\
 \hspace{-0.3cm} &&=\ln(2) f \vol+\frac{3\ln(1-f)}{2}-\frac{2{\spin_A}^2}{(1-f)\vol}-\frac{1}{2}\delta_{f,\frac{1}{2}} +o(1). \nonumber
\end{eqnarray}
Note that it is interesting that in the expansion of $\varphi({\spin_A})$ the term $\ln(1+2{\spin_A})$ is exactly canceled by a similar term appearing in $S_{\mathrm{Page}}(n_{\spin_A}^A,n_{\spin_A}^B)$, so that there is no $\ln(\vol)$ term at $\spin=0$.

For large $\vol$, we can evaluate the sum as an integral
\begin{align}
    \braket{\ent_A}_{\substack{\spin=0,\\ \spin_z=0}}=\sum\rho({\spin_A})\varphi({\spin_A}) = \int \rho({\spin_A})\varphi({\spin_A})d{\spin_A} +o(1),
\end{align}
and then do a rescaling by introducing $j_A={\spin_A}/\sqrt{\vol}$, so that $\braket{\ent_A}_{\spin=0,\spin_z=0}=\int_0^\infty \sqrt{\vol}\rho(j_A)\varphi(j_A)dj_A$. The leading orders terms for $f=\vol_A/\vol\leq \frac12$ read
\begin{equation}\label{eq:S=0_Asymptotic}
    \braket{\ent_A}_{\substack{\spin=0,\\ \spin_z=0}}=\ln(2)f\vol+\frac{3[f+\ln(1-f)]}{2} -\frac{1}{2}\delta_{f,\frac{1}{2}}+o(1)\,,
\end{equation}
where, as before, $o(1)$ indicates corrections that vanish in the thermodynamic limit. The first two terms in Eq.~\eqref{eq:S=0_Asymptotic} were obtained for a related problem away from $f=\frac12$ in Ref.~\cite{majidy_non-abelian_2023}. 

The exact result in Eq.~\eqref{eq:S=0_Asymptotic} has some important properties that we would like to emphasize: (i) The leading volume-law term has the expected maximal coefficient advanced by Eq.~\eqref{eq:AsymptoticHilbSpcDim} at $\spin=0$. (ii) The first subleading correction is $O(1)$. It has the same structure as the one in the presence of U(1) symmetry [see Eq.~\eqref{eq:n_lead}, for which $n=\frac12$ is equivalent to $\spin_z=0$ here], with one term that is a function of $f$ and a $-\frac12$ that appears only at $f=\frac12$. Note that the prefactor of the function of $f$ is different in Eqs.~\eqref{eq:S=0_Asymptotic} and~\eqref{eq:n_lead}. 

Since all other sectors with $\spin=O(1)$ have $\sd\rightarrow0$ as $\vol\rightarrow\infty$, it is to be expected given Eq.~\eqref{eq:n_lead} that all those sectors will exhibit the same leading volume-law term. This and the nature of the first subleading correction for $\spin=O(1)$ are explored next.

\subsection{Spin $\spin=O(1)$}

When $\spin\neq 0$, random states cannot be decomposed into direct sums of tensor products because there are many possible pairings between $\spin_A$ [Eq.~\eqref{eq:rangesJA}] and $\spin_B$ [Eq.~\eqref{eq:rangesJB}]. A basis $\{\ket{\phi_{ab}}\}$ of $\mathcal{H}_{\spin,\spin_z=0}$, in terms of $\ket{\spin_A,m_A}\otimes\ket{\spin_B,m_B}$, can be written as
\begin{align}
\ket{\phi_{ab}}=\hspace{-0.2cm}\sum^{\min[\spin_A,\spin_B]}_{m=-\min[\spin_A,\spin_B]}\hspace{-0.5cm}c_m(\spin,\spin_A,\spin_B)\ket{\spin_A,m}_a\otimes\ket{\spin_B,-m}_b\,,
\end{align}
with $c_m(\spin,\spin_A,\spin_B)=\braket{\spin_A,m,\spin_B,-m|\spin,0}$ being the corresponding CG coefficient. 

A random state $\ket{\phi}$ in $\mathcal{H}_{\spin,\spin_z=0}$, and its reduced density matrix $\hat {\rho}_A=\mathrm{Tr}\ket{\phi}\bra{\phi}$, therefore take the form
\begin{widetext}
\begin{eqnarray}
    \ket{\phi}\hspace{-0.1cm} &=&\hspace{-0.2cm}\sum^{\min[\frac{\vol_A}{2},\spin+\frac{\vol_B}{2}]}_{\spin_A=\max[\spin_{\min},\spin-\frac{\vol_B}{2}]}\sum^{\min[\frac{\vol_B}{2},\spin+\spin_A]}_{\spin_B=\max[\spin_{\min},|\spin-\spin_A|]} \sum^{n^A_{\spin_A}[\vol_A]}_{a=1} \sum^{n^B_{\spin_B}[\vol_B]}_{b=1} \sum^{\min[\spin_A,\spin_B]}_{m=-\min[\spin_A,\spin_B]} \hspace{-0.1cm} W^{\spin_A\spin_B}_{ab}c_m(\spin,\spin_A,J_B)\ket{\spin_A,m}_a\otimes\ket{\spin_B,-m}_b,\nonumber \\ \label{eq:ran-psi}\\
    \hat {\rho}_A \hspace{-0.1cm} &=&\hspace{-0.2cm}\sum^{\min[\frac{\vol_A}{2},\spin+\frac{\vol_B}{2}]}_{\spin_A,\spin'_A=\max[\spin_{\min},\spin-\frac{\vol_B}{2}]}\sum^{\min[\frac{\vol_B}{2},\spin+\spin_A,\spin+\spin_A']}_{\spin_B=\max[\spin_{\min},|\spin-\spin_A|,|\spin-\spin_A'|]}\sum^{n^A_{\spin_A}[\vol_A]}_{a=1}\sum^{n^A_{\spin'_A}[\vol_A]}_{a'=1}\sum^{n^B_{\spin_B}[\vol_B]}_{b=1}\sum^{\min[\spin_A,\spin_A',\spin_B]}_{m=-\min[\spin_A,\spin_A',\spin_B]} W^{J_AJ_B}_{ab}(W^{\spin_A'\spin_B}_{a'b})^* \nonumber \\
    &&\hspace{8.5cm} \times\,  c_m(\spin,\spin_A,\spin_B)c^*_m(\spin,\spin_A',\spin_B)\ket{\spin_A,m}_a\bra{\spin_A',m}_{a'},\label{eq:ran-rho_A}
\end{eqnarray}
\end{widetext}
where $W^{\spin_A\spin_B}_{ab}\in \mathbb{C}$ are Gaussian random variables with zero mean and fixed variance drawn from the fixed trace ensemble, \ie the normalization of the state $\ket{\phi}$ requires 
\begin{equation}
    \sum_{\spin_A,\spin_B}\sum^{n^A_{\spin_A}[\vol_A]}_{a=1} \sum^{n^B_{\spin_B}[\vol_B]}_{b=1}|W^{\spin_A\spin_B}_{ab}|^2=1,
\end{equation}
where the limits of the sums over $\spin_A$ and $\spin_B$ are given by Eqs.~\eqref{eq:rangesJA} and~\eqref{eq:rangesJB}.

One can see that the matrix $\hat{\rho}_A$ is block-diagonal with respect to the spin component $m$ in the subsystem $A$, but in principle has ``interferences'' between different $\spin_A$ and $\spin_A'$. Only in the special case in which $\spin=0$ we effectively have $\delta_{\spin_A,\spin_A'}\delta_{\spin_A,\spin_B}$, which leads to the block structure over $\spin_A$ discussed for $\spin=0$. If $\spin$ is not extensive ($\sd=0$ in the limit $\vol\rightarrow\infty$), \ie $\spin=O(1)$ or $\spin=O(\sqrt{\vol})$, we expect that the entries of $\hat{\rho}_A$ have a band structure around $\spin_A=\spin_A'$.

\begin{figure}[!t]
    \includegraphics[width=0.98\columnwidth]{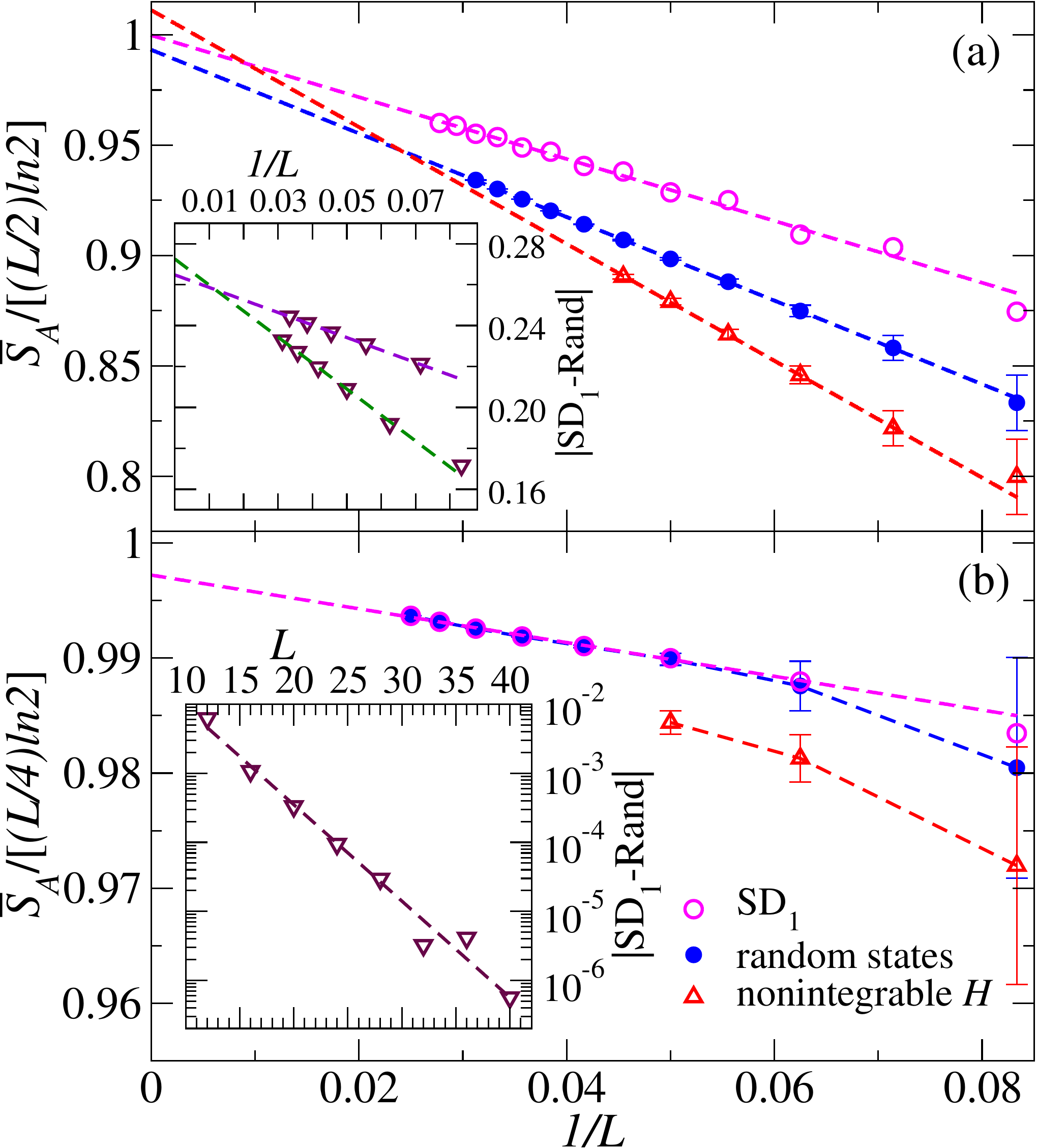}
    \vspace{-0.2cm}
    \caption{{\it Scaling of $\bar \ent_A$ for $\spin=1$.} $\bar \ent_A$ vs $1/\vol$ for eigenstates of the quantum-chaotic Hamiltonian~\eqref{eq:Hamiltonian Spin 1/2}, random pure states, and the SD$_\text{1}$, at $f=\frac12$ (a) and at $f=\frac14$ (b). The error bars show the standard deviation of the averages, and the dashed lines show linear ($a+b/\vol$) fits to all data sets in (a) and to the SD$_\text{1}$ results in (b). (Insets) Relative differences between the random states and the SD$_\text{1}$ results, which are consistent with $1/\vol$ and $e^{-a\vol}$ decays in (a) and (b), dashed lines, respectively. The results for systems sizes that are even or odd multiples of $2$ in (a) have a different slope in their finite-size scalings, but approach a similar $O(1)$ number ($\sim0.27$) as $1/\vol\rightarrow0$. Fits ($a+b/\vol$) in that case are carried out using the results for the largest $4$ system sizes in each case.}
    \label{fig:J=1}
\end{figure}

\begin{figure}[!t]
    \centering
    \includegraphics[width=0.98\columnwidth]{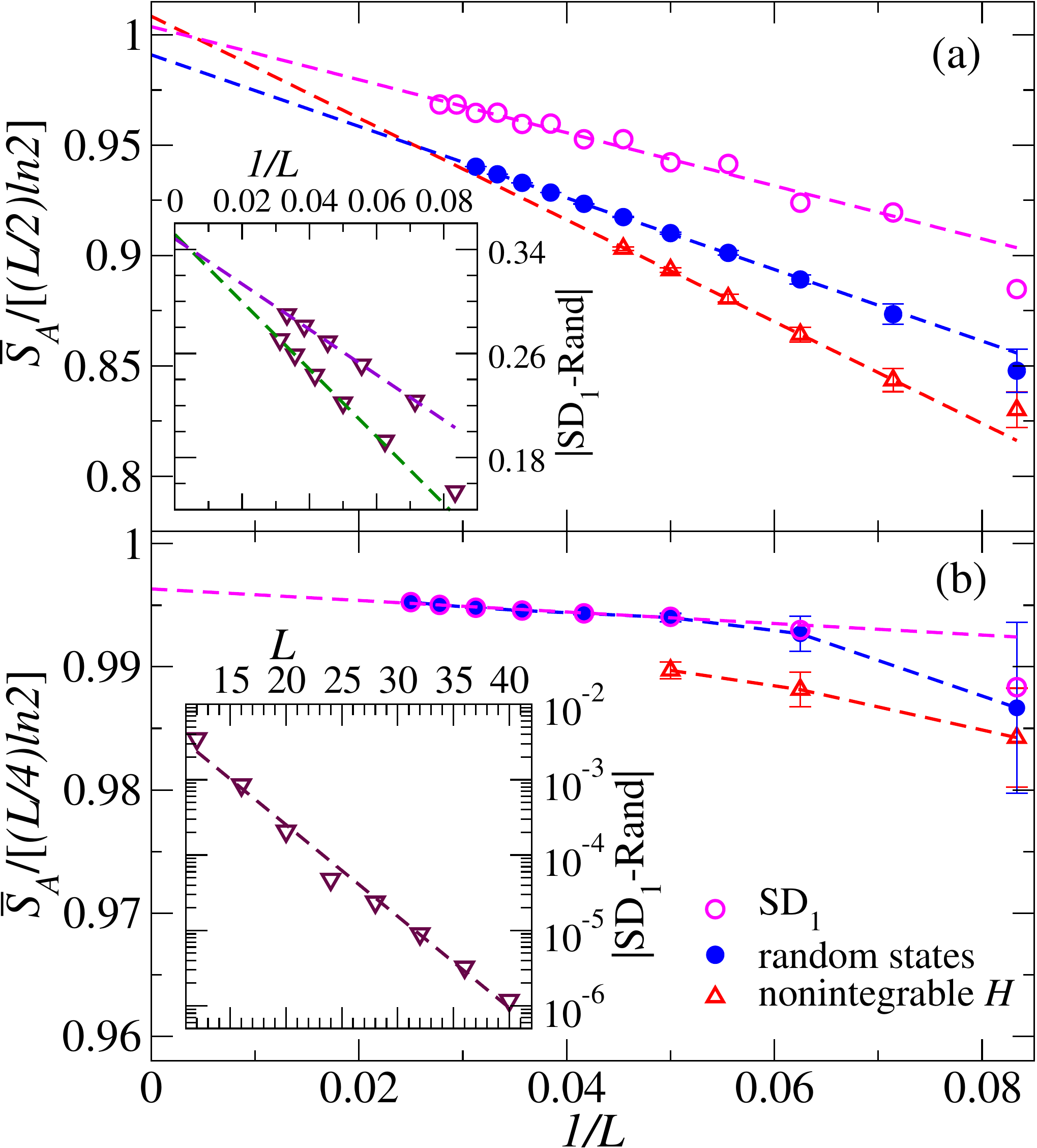}
    \vspace{-0.2cm}
    \caption{{\it Scaling of $\bar \ent_A$ for $\spin=2$.} Same as Fig.~\ref{fig:J=1} but for $\spin=2$. In the inset in (a), the fits cross the $y$ axis at $\sim0.35$.}
    \label{fig:J=2}
\end{figure}

In Figs.~\ref{fig:J=1} and~\ref{fig:J=2}, we plot numerical results for the average entanglement entropy obtained for random pure states using $\hat {\rho}_A$ in Eq.~\eqref{eq:ran-rho_A} for $\spin=1$ and $2$, respectively, at $f=\frac12$ (a) and $\frac14$ (b). [We use real coefficients in the evaluation of Eq.~\eqref{eq:ran-rho_A}, see Appendix~\ref{sec:randcoeff}.] The results in the plots are normalized by the expected leading volume-law term. We also plot in Figs.~\ref{fig:J=1} and~\ref{fig:J=2} numerical results for the average entanglement entropy of highly excited eigenstates of the quantum-chaotic (nonintegrable) Hamiltonian [Eq.~\eqref{eq:Hamiltonian Spin 1/2} with $\cnn=3$]. For random pure states for both values of $J$ and $f$ shown, and for Hamiltonian eigenstates for both values of $J$ shown at $f=\frac12$ (at $f=\frac14$ we have insufficient data points), the numerical results are consistent with the leading volume-law term in $\bar\ent_A$ being the expected maximal result (the $y$-axis intercept at $1/\vol=0$ is close to 1), and with the first subleading correction being $O(1)$ (the numerical results follow linear $a+b/\vol$ fits). Those results suggest that Eq.~\eqref{eq:S=0_Asymptotic} applies for $\spin=O(1)>0$, but with an $O(1)$ correction that depends on $J$.

We note that Eq.~\eqref{eq:ran-rho_A} allows us to numerically compute the entanglement entropy averages over random pure states for larger system sizes than those accessible by the calculation involving Eq.~\eqref{eq:randavenum}, which requires generating an exponentially large basis $\{\ket{\spin,0}_i\}_{i=1}^{n_\spin}$ for $\mathcal{H}_\spin$. In order to carry out numerical calculations for random pure states in even larger system sizes, as well as to make analytic progress later for $\spin=O(\vol)$, we introduce an approximation to evaluate Eq.~\eqref{eq:ran-rho_A} that is motivated by the $\spin=0$ case. We call this approximation the ``spin decomposition 1'', in short SD$_\text{1}$. The SD$_\text{1}$ approximation ignores the ``interference'' between different $\spin_A$ and $\spin_A'$, \ie it assumes that $\hat{\rho}_A$ is also block diagonal with respect to $J_A$. This means that we include a Kronecker delta $\delta_{\spin_A,\spin_A'}$ in the sum in Eq.~\eqref{eq:ran-rho_A}. 

The corresponding Hilbert space decomposition resembles Eq.~\eqref{eq:HJ0Jz0} and is given by
\begin{align}
\mathcal{H}^{\mathrm{SD}_1}_{\substack{\spin,\spin_z=0}}=\bigoplus^{\min[\frac{\vol_A}{2},J+\frac{\vol_B}{2}]}_{\spin_A=\max[\spin_{\min},J-\frac{\vol_B}{2}]} \mathcal{H}^{\spin}_{\spin_A}\,,
\end{align}
where $\mathcal{H}^{\spin}_{\spin_A}\subset \mathcal{H}_{\spin_A}\otimes \mathcal{H}_{B,\spin_A}$ contains the $d_{\spin_A}=n_{\spin_A}^A\times n_{B,J_A}^B$ states that have fixed spin $\spin_A$ in subsystem $A$ and total spin $\spin$. Here, $\mathcal{H}_{B,\spin_A}$ is a direct sum over all $\spin_B$ Hilbert spaces that can combine with $J_A$ to give total spin $J$ and their number is given by
\begin{align}
    n_{B,J_A}^B=\sum^{\min[\frac{\vol_B}{2},\spin+\spin_A]}_{\spin_B=\max[\spin_{\min},|\spin-\spin_A|]}n^{B}_{J_B}\,.
\end{align}
The average entanglement entropy can be obtained by computing the (Haar random) average entanglement entropy $\braket{{S_A}}^J_{\spin_A}$ over the restricted subspace $\mathcal{H}^{\spin}_{\spin_A}$ using the equivalent of Eq.~\eqref{eq:EntCG+Page}, plugging it into Eq.~\eqref{eq:tensor-structure} instead of $\braket{{S_A}}^0_{\spin_A}$, and using the appropriate dimensions $d_{\spin_A}=n_{\spin_A}^A\times n_{B,J_A}^B$ and $d=\sum_{\spin_A} d_{\spin_A}$.

In Figs.~\ref{fig:J=1} and~\ref{fig:J=2}, we plot numerical results for the average entanglement entropy obtained for random pure states using the SD$_\text{1}$ for $\spin=1$ and $2$, respectively, at $f=\frac12$ (a) and $\frac14$ (b). (We use real coefficients in the evaluation of the SD$_\text{1}$, see Appendix~\ref{sec:randcoeff}.) For both values of $J$, one can see that as the system size increases the SD$_\text{1}$ results at $f=\frac14$ become indistinguishable from the numerical evaluation of Eq.~\eqref{eq:ran-rho_A}. At $f=\frac12$, on the other hand, the SD$_\text{1}$ results are always greater than those obtained using Eq.~\eqref{eq:ran-rho_A}, but the difference appears to be $O(1)$ because the linear fits intercept the $y$ axes at points close to 1, only the slopes are different. We show results for the scaling of the differences between the numerical calculation for random pure states and the SD$_\text{1}$ in the insets in Figs.~\ref{fig:J=1} and~\ref{fig:J=2}. They make apparent that at $f=\frac14$ (and we expect the same for other values of $f\neq \frac12$) the differences vanish exponentially with increasing system size. At $f=\frac12$, on the other hand, the differences appear to converge to small $O(1)$ numbers, $\sim 0.27$ for $\spin=1$ and $\sim 0.35$ for $\spin=2$, in the thermodynamic limit. 

Summarizing our results for $\spin=O(1)$, we provided numerical evidence that Eq.~\eqref{eq:S=0_Asymptotic} applies for $\spin=O(1)>0$, but with an $O(1)$ correction that depends on $J$. Furthermore, our results in the insets in Figs.~\ref{fig:J=1} and~\ref{fig:J=2} show that at $f\neq \frac12$ using the SD$_\text{1}$ introduces an exponentially small error when evaluating the average entanglement entropy for $\spin=O(1)$, while the error appears to be an $O(1)$ error at $f=\frac12$. 

Since $\sd=0$ in the thermodynamic limit also for $\spin=O(\sqrt{\vol})$, we expect the leading volume-law term in Eq.~\eqref{eq:S=0_Asymptotic} to also apply to that case. The nature of the subleading corrections for $\spin=O(\sqrt{\vol})$ is something that will need to be studied in future works.

\subsection{Spin $\spin=O(\vol)$} 

We conclude our study of the average entanglement entropy of random pure states by considering the case $\spin=O(\vol)$. As discussed in Sec.~\ref{sec:hilbspacdim}, in this case the ratio $n_\spin/D$ decays exponentially with $\vol$.

We consider first $\spin=\vol/2$, which is the largest spin. The sector $\mathcal{H}_{\spin=\vol/2,\spin_z=0}$ contains only one state, with $\spin_A=\vol_A/2=f\spin$ and $\spin_B=\vol_B/2=(1-f)\spin$, such that
\begin{align}
    \ket{\psi}=\sum^{\min[\spin_A,\spin_B]}_{m=-\min[\spin_A,\spin_B]}c_m(\spin,\spin_A,\spin_B)\ket{\spin_A,m}\otimes\ket{\spin_B,-m}\,,
\end{align}
where $c_m(\spin,\spin_A,\spin_B)$ is the CG coefficient: 
\begin{eqnarray}
    &&\braket{J_A,m,J_B,-m|\spin=J_A+J_B,0}\nonumber\\&&= (-1)^{2 \spin_A-2 \spin_B}\, \sqrt[4]{\pi }\, \sqrt{2 \left(\spin_A+\spin_B\right)+1}\,\times\\&& \quad  \sqrt{\tfrac{\left(2 \spin_A\right)! \left(2 \spin_B\right)! 2^{-2 \spin_A-2 \spin_B} \left(\spin_A+\spin_B\right)!}{2\left(\spin_A+\spin_B+\frac{1}{2}\right)! \left(\spin_A-m\right)! \left(\spin_A+m\right)! \left(\spin_B-m\right)! \left(\spin_B+m\right)!}} \,. \nonumber
\end{eqnarray}
This state has a simpler form when written in terms of the tensor product basis of individual $\hat{\spin}^z_i$, that is, it is a uniform superposition of all the base states with zero total magnetization $\spin^z=0$ given by~\cite{salerno_05}
\begin{equation}
    \ket{\psi} \sim (\hat{\spin}^+)^{\frac{\vol}{2}} \bigotimes_{i=1}^\vol \ket{\spin^z_i=-\frac12}\,,
\end{equation}
where $\hat{\spin}^+=\sum_{i=1}^\vol \hat{\spin}^+_i$ with the raising operators $\hat{\spin}^+_i=\hat{\spin}^x_i+i\hat{\spin}^y_i$ at site $i$.  

The reduced density operator becomes
\begin{align}
    \hat \rho_A=\sum^{\min[\spin_A,\spin_B]}_{m=-\min[\spin_A,\spin_B]}|c_m(\spin,\spin_A,\spin_B)|^ 2\ket{\spin_A,m}\otimes\bra{\spin_A,m},
\end{align}
and the entanglement entropy $\braket{\ent_A}_{\spin=\frac{\vol}{2},\spin_z=0}=\ent_A(\ket{\psi})$ is thus the one of the CG coefficients (as probability distribution). In the limit of $\vol\to\infty$ for fixed $0<f<\frac12$, the distribution of $|c_m(\spin,\spin_A,\spin_B)|^2$ in $m$ approaches a normal distribution with average $m=0$ and standard deviation $\sqrt{\frac{f(1-f)\vol}{4}}$. A closed form for the leading term in the entanglement entropy of this state can be obtained using the distribution of the CG coefficients $c_m(\spin,\spin_A,\spin_B)=\braket{\spin_A,m;\spin_B,-m|\spin,\spin_z=0}$~\cite{salerno_05}, so that for $0<f<1$,
\begin{equation}
    \braket{\ent_A}_{\substack{\spin=\frac{\vol}{2},\\ \spin_z=0}}=\frac{1}{2}\ln\left[\frac{\pi e f(1-f) \vol}{2}\right]+o(1).\label{eq:J=L/2}
\end{equation}

To make analytic progress for $\spin=O(\vol)<\vol/2$, we introduce a ``spin decomposition 2'' (SD$_\text{2}$) with an extra simplifying assumption on top of the SD$_\text{1}$ discussed for $\spin=O(1)>0$. In the SD$_\text{2}$, we assume that the leading contributions to the entanglement entropy come from the terms in $\hat{\rho}_A$ [Eq.~\eqref{eq:ran-rho_A}] where $J_B=J-J_A$, which amounts to including a product of Kronecker deltas $\delta_{\spin_B,\spin-\spin_A}\delta_{\spin_A,\spin_A'}$ in the sum in Eq.~\eqref{eq:ran-rho_A}. This assumption is justified by the observation that for large $L$ and fixed $J_A<J$, the number $n^B_{J_B}$ falls off exponentially as we increase $J_B$ from $J_B=J-J_A$, \ie most of the states with fixed $J$, $J_A$, and $J_B$ satisfy the relation $\spin=J_A+J_B$ (the SD$_\text{2}$ is exact for $\spin=\vol/2$, for which there is only one state). For the SD$_\text{2}$, we thus only compute the average entanglement entropy over those states.

\begin{figure*}[!t]
    \centering
    \includegraphics[width=.95\textwidth]{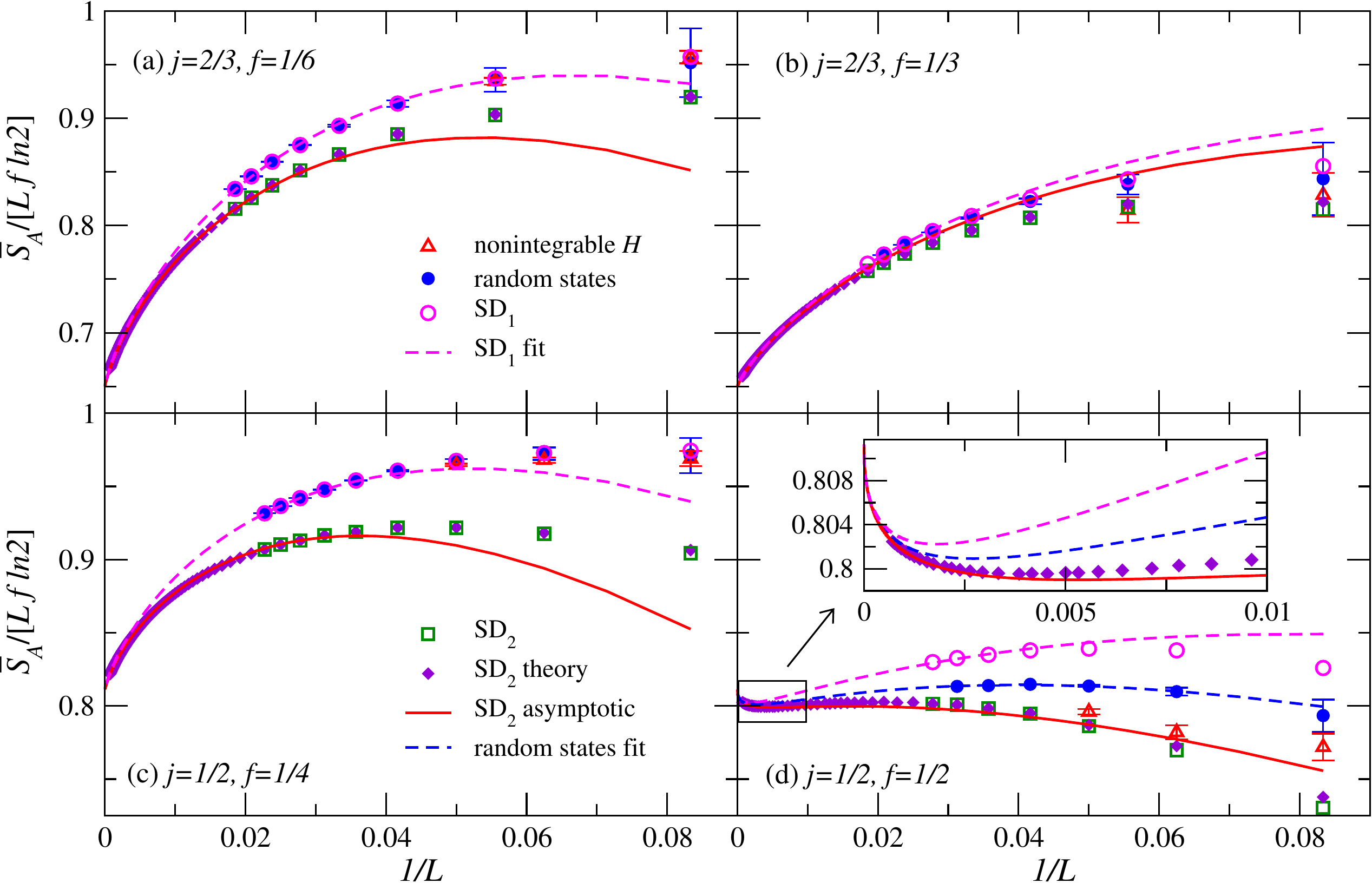}
    \caption{{\it Scaling of $\bar \ent_A$ for $\spin=O(\vol)$.} $\bar \ent_A$ vs $1/\vol$ for eigenstates of the quantum-chaotic Hamiltonian~\eqref{eq:Hamiltonian Spin 1/2}, random pure states, the SD$_\text{1}$, and the SD$_\text{2}$ [Eq.~\eqref{eq:SA-SD_B} and numerical results], at $\spin=\vol/3$ [(a) for $f=\frac16$ and (b) for $f=\frac13$] and $\spin=\vol/4$ [(c) for $f=\frac14$ and (d) for $f=\frac12$]. The inset in (d) is a zoom into the $1/\vol\rightarrow 0$ regime. The solid lines show the predictions of Eq.~\eqref{eq:EntanglementJ=O(L)}, while the dashed lines are fits of the results (for the largest 4 values of $L$) of the SD$_\text{1}$, and in (d) also of the random pure state averages, to Eq.~\eqref{eq:EntanglementJ=O(L)} plus an $O(1)$ constant as the only fitting parameter.}
    \label{fig:J=O(L)}
\end{figure*}

This yields the Hilbert space
\begin{align}
\mathcal{H}^{\mathrm{SD}_2}_{\substack{\spin,\spin_z=0}}=\bigoplus^{\min[\frac{\vol_A}{2},J+\frac{\vol_B}{2}]}_{\spin_A=\max[\spin_{\min},J-\frac{L_B}{2}]} \mathcal{H}^{\spin}_{\spin_A,\spin-\spin_A}\,,
\end{align}
where $\mathcal{H}^{\spin}_{\spin_A,\spin-\spin_A}\subset \mathcal{H}_{J_A}\otimes\mathcal{H}_{J-J_A}$ contains the $d_{J_A}=n^A_{J_A}\times n^B_{J-J_A}$ states with fixed $J_A$, $J_B=J-J_A$, and $J$.

The resulting density matrix $\hat{\rho}_A$ of a Haar random state is thus block diagonal over both $J_A$ and $m$ [similar to Eq.~\eqref{eq:rho_A J=0}] with blocks given by
\begin{align}
    W^{J_A,J-J_A}(W^{J_A,J-J_A})^\dagger |c_m(J,J_A,J-J_A)|^2\,.\label{eq:JA-J-JA-block}
\end{align}
The normalization of the original state is then equivalent of requiring
\begin{align}
    \sum^{\min[\frac{\vol_A}{2},\spin+\frac{\vol_B}{2}]}_{\spin_A=\max[\spin_{\min},\spin-\frac{\vol_B}{2}]}\mathrm{Tr}(W^{J_A,J-J_A}{W^{J_A,J-J_A}}^\dagger)=1\,,
\end{align}
\ie it is equivalent to setting $W^{J_A,J_B}=0$ in Eq.~\eqref{eq:ran-psi} for $J_B\neq J-J_A$.

We can compute the (Haar random) average entanglement entropy $\braket{{S_A}}^J_{\spin_A,\spin-\spin_A}$ over the restricted subspace $\mathcal{H}^{\spin}_{\spin_A,\spin-\spin_A}$ analytically, as it is the entropy associated to a block of the form in Eq.~\eqref{eq:JA-J-JA-block}, though with the simpler constraint 
\begin{equation}
\mathrm{Tr}(W^{J_A,J-J_A}{W^{J_A,J-J_A}}^\dagger)=1,
\end{equation}
as we only sample in the respective block. The resulting entropy can be computed in full analogy to Eq.~\eqref{eq:EntCG+Page}:
\begin{eqnarray}\label{eq:SA sum SD2}
    &&\hspace{-1cm}\braket{\ent_A}_{\spin_A,\spin-\spin_A}^\spin=\ent_{\mathrm{CG}}(\spin_A) + \ent_{\mathrm{Page}}(n^A_{\spin_A},n^B_{J-\spin_A})\,.
\end{eqnarray}
The average entanglement entropy is then
\begin{align}
        \braket{S_A}^{\text{SD}_\text{2}}_{\substack{\spin,\quad\\ \spin_z=0}}\hspace{-0.2cm} =\! \sum_{\spin_A} \frac{d_{\spin_A}}{d} \left[ \braket{{S_A}}^J_{\spin_A,\spin-\spin_A}\!\!+\!\Psi(d\!+\!1)\! -\! \Psi(d_{\spin_A}\!+\!1)\right],\label{eq:SA-SD_B}
\end{align}
with $d_{\spin_A}=n^A_{J_A}\times n^B_{J-J_A}$, $d=\sum_{J_A}d_{\spin_A}$. The leading-order terms for large $L$ are given by (see Appendix~\ref{sec:S3})
{\setlength\arraycolsep{0.7pt}
\begin{eqnarray}
    &&\braket{\ent_A}^{\text{SD}_\text{2}}_{\substack{\spin=O(\vol),\\ \hspace{-0.45cm} \spin_z=0}}= s_A(j)\, f\, \vol  + \frac{\sqrt{1-j^2}\,\ln\left(\frac{1-j}{1+j}\right)}{2\sqrt{2\pi}}\,\sqrt{\vol}\,\delta_{f,1/2} \nonumber \\ && \hspace{2.25cm} +\braket{\ent_A}_{\substack{\spin=\frac{\vol}{2},\\ \spin_z=0}} + h(j,f) + o(1), \label{eq:EntanglementJ=O(L)}\\
    && s_A(j) = - \left[\frac{1+j}{2} \ln \left(\frac{1+j}{2}\right)  +  \frac{1-j}{2}  \ln  \left(\frac{1-j}{2}\right)\right]\!, \label{eq:CoefficientJ=O(L)}\\
    &&h(j,f) = \ln\left(\frac{2j^{3/2}}{\sqrt{1-j^2}}\right)-\frac{1-2f(1-j)}{2j}\ln\left(\frac{1+j}{1-j}\right)\, \nonumber\\
    &&\hspace{1.4cm}+(1-\delta_{j,1}) \frac{f+\ln(1-f)}{2}  .\label{eq:g-formula}
\end{eqnarray}
}Three points to emphasize about $\braket{\ent_A}^{\text{SD}_\text{2}}_{\spin=O(\vol),\spin_z=0}$ in Eq.~\eqref{eq:EntanglementJ=O(L)} are as follows: (i) The coefficient of the volume in the leading term is the one advanced in Eq.~\eqref{eq:AsymptoticHilbSpcDim}. (ii) There is a $\sqrt{\vol}$ correction that appears at $f=\frac12$ when $j\neq0,1$. (iii) The subleading $\ln L$ correction for $j<1$ becomes the leading term at $J=L/2$.

In Figs.~\ref{fig:J=O(L)}(a)--\ref{fig:J=O(L)}(c), we show results for $\bar \ent_A$ for eigenstates of the quantum-chaotic Hamiltonian [Eq.~\eqref{eq:Hamiltonian Spin 1/2} with $\cnn=3$], random pure states, and the SD$_\text{1}$, at $\spin=\vol/3$ [(a) for $f=\frac16$ and (b) for $f=\frac13$] and at $\spin=\vol/4$ [(c) for $f=\frac14$], all away from $f=\frac12$. As in Figs.~\ref{fig:J=1}(b) and \ref{fig:J=2}(b), the random states and the SD$_\text{1}$ results become indistinguishable as $L$ increases (because their difference is exponentially small in $L$), and the Hamiltonian eigenstates results are very close to them. The SD$_\text{2}$ results are in all cases smaller, but they approach the others with increasing $L$. The differences between the SD$_\text{1}$ and SD$_\text{2}$ results are consistent with the SD$_\text{2}$ approximation introducing an $O(1)$ error. This expectation is supported by the fact that in Figs.~\ref{fig:J=O(L)}(a)--\ref{fig:J=O(L)}(c) we show that the same equation that describes the SD$_\text{2}$ results with increasing $L$ [Eq.~\eqref{eq:EntanglementJ=O(L)}], describes the SD$_\text{1}$ results for the largest system sizes after we add an $O(1)$ constant to Eq.~\eqref{eq:EntanglementJ=O(L)} as a fitting parameter. The same applies to the SD$_\text{1}$ and random state results in Fig.~\ref{fig:J=O(L)}(d) at $\spin=\vol/4$ and $f=\frac12$. 

At $f=\frac12$, the SD$_\text{1}$ already introduces an $O(1)$ error, so the SD$_\text{1}$ results are visibly greater from those obtained using the full reduced density matrix for random pure states. The SD$_\text{2}$ results, on the other hand, are smaller than those for random pure states. With increasing system size, all the numerical results in Fig.~\ref{fig:J=O(L)}(d) approach each other, which suggests that the leading volume-law term is the same for all calculations. Having both a $\ln \vol$ and a $\sqrt{\vol}$ correction at $\spin=\vol/4$ and $f=\frac12$ produces finite-size effects that are non-monotonic as $\vol\rightarrow\infty$. The inset highlights the regime in which the $\sqrt{\vol}$ term becomes the dominant subleading correction in SD$_\text{2}$.

\section{Summary and discussion}\label{sec:summary}

We studied the effect of the SU(2) symmetry in the average entanglement entropy of highly excited Hamiltonian eigenstates of spin $\ms=\frac12$ and 1 models in the $\spin_z=0$ subspace in sectors with different fixed spin $\spin$. Our numerical results provide evidence that the leading volume-law term in the average entanglement entropy of highly excited eigenstates of quantum-chaotic (integrable) Hamiltonians is the same as (different from) that obtained from maximally mixed states in the appropriate sectors of the Hilbert space of subsystem A. We also carried out analytical and numerical calculations for random pure states with spin $\ms=\frac12$. Our results indicate (prove for $J=0$) that the leading term in the average entanglement entropy of random pure states is also the one predicted by the maximally mixed state in the appropriate sector of the Hilbert space of subsystem A, as we find for the average entanglement entropy of highly excited eigenstates of quantum-chaotic Hamiltonians. Hence, our results suggest that the average entanglement entropy can be used as a diagnostic of quantum chaos and integrability in models with non-Abelian symmetries. 

More specifically, for $\ms=\frac12$ in sectors where $\spin=O(1)$, whose dimension $n_\spin$ divided by the dimension $D$ of the $\spin_z=0$ subspace vanishes as $n_\spin/D\propto 1/L$, our results indicate (prove for the average over random states with $\spin=0$) that the leading volume term in the average entanglement entropy is maximal (identical to that of the average over random states in the $\spin_z=0$ subspace), while the first subleading correction is $O(1)$. We advance that the same is true about the leading term of the larger $\spin=O(\sqrt{\vol})$ sectors, for which $n_\spin/D\propto 1/\sqrt{\vol}$. A direct study of those sectors remains a challenge for future analytical and numerical studies. 

We find that the SU(2) symmetry plays its most distinctive role in sectors with $\spin=O(\vol)$, for which $n_\spin/D$ vanishes exponentially with increasing system size. Using a spin decomposition (SD$_\text{2}$) supplemented by numerical results for random pure states with $\ms=\frac12$, we showed that in the $\spin=O(\vol)$ sectors the coefficient $s_A$ of the leading volume-law term depends on the spin density $\sd=2\spin/\vol$, with $s_A(\sd \rightarrow 0)=\ln 2$ and $s_A(\sd \rightarrow 1)=0$ [see Eq.~\eqref{eq:CoefficientJ=O(L)}]. Away from $f=\frac12$, we find the first subleading correction to be $\ln \vol$ (this correction becomes the leading term at $\sd=1$). Subleading corrections of this form do not appear in the presence of U(1) symmetry [see Eq.~\eqref{eq:n_lead}], and they may be a hallmark of non-Abelian symmetries. Furthermore, at $f=\frac12$ and $\sd\neq 1$, we found that the first subleading correction is $\propto \sqrt{\vol}$. 

Our numerical results indicate that Eq.~\eqref{eq:EntanglementJ=O(L)}, which is one of the main analytical results of this work, differs from the exact Haar-random average in the $O(1)$ correction. A challenging task that we plan to tackle next is computing the exact value of the $O(1)$ correction for the Haar-random average. As a first step to achieve this, we intend to compute the equivalent of Eq.~\eqref{eq:EntanglementJ=O(L)} in the context of the SD$_\text{1}$, which our numerical results indicate approaches the exact result for $f\neq \frac12$ exponentially fast with increasing $\vol$. Another interesting question that we plan to explore is the effect of non-Abelian symmetries in the symmetry-resolved entanglement entropy. The effect of the Abelian U(1) symmetry in the symmetry-resolved entanglement entropy was recently studied in Ref.~\cite{murciano_23}.

\emph{Acknowledgments}.--- We acknowledge the support of the National Science Foundation, Grants No.~PHY-2012145 and No.~PHY-2012145 (R.P. and M.R.), Grant No.~DMR-1851987 (G.R.F.), of the John Templeton Foundation, Grant No.~62312 (L.H.) as part of the \href{https://www.templeton.org/grant/the-quantuminformation-structure-ofspacetime-qiss-second-phase}{``The Quantum Information Structure of Spacetime'' Project (QISS)}, and of the Alexander von Humboldt Foundation (L.H.). The computations were done in the Institute for Computational and Data Sciences’ Roar supercomputer at Penn State. The opinions expressed in this publication are those of the authors and do not necessarily reflect the views of the John Templeton Foundation.

\appendix

\section{Hilbert space dimensions for spin $\ms$}\label{sec:AppHSD}

Let us briefly review how to compute the multiplicity $n^{\ms}_\spin$ introduced in Eq.~\eqref{eq:H-decomposition} using group theory~\cite{hall2013lie}.

The \emph{character} of a group element $g\in\mathcal{G}$ in a representation $\rho: \mathcal{G}\to \mathrm{Lin}(\mathcal{H}^\ms)$ is given by the trace function $\chi(g)=\mathrm{Tr}\rho(g)$. In the case of the spin-$J$ representation of $\mathrm{SU}(2)$, we can use the \emph{Weyl character formula}:
\begin{align}
    \chi_J(g)=\frac{\sin(2J+1)\theta}{\sin{\theta}}\,,
\end{align}
where the group element is parametrized by a coordinate $\theta\in[0,2\pi]$ and two other coordinates, which $\chi_J(g)$ does not depend on. The (invariant) Haar measure after integrating out the other two coordinates is given by $(\sin^2\theta/\pi) d\theta$.

A key property of characters is that they multiply when taking tensor products of representations and add up when taking direct sums of representations. Moreover, the character functions (on compact groups) are orthonormal with respect to the normalized Haar measure. Therefore, to determine how often the representation $J$ appears in the tensor product $\ms^{\otimes L}$, one can just evaluate the integral
\begin{align}
    n^\ms_J\equiv n^\ms_J(\vol)=\frac{1}{\pi}\int \chi_J(\theta)\chi_{\ms}(\theta)^L \sin^2{\theta}d\theta\,,
\end{align}
where $\chi_J(\theta)$ refers to the spin-$J$ representation we want to count and $\chi_\ms(\theta)^L$ is the character of the tensor product representation $\ms^{\otimes L}$.

We can rewrite this integral using $z=e^{-i\theta}$ as 
\begin{align}
    n^\ms_J&=i\oint\tfrac{(z^2-1)(z^{1+2J}-z^{-(1+2J)})}{4\pi}\left(\tfrac{z^{1+2\ms}-z^{-(1+2\ms)}}{z-z^{-1}}\right)^Ldz \nonumber \\
    &=i\oint\tfrac{(z^2-1)z^{1+2J}}{2\pi}\left(\tfrac{z^{1+2\ms}-z^{-(1+2\ms)}}{z-z^{-1}}\right)^Ldz,
\end{align}
where the contour integral follows the unit circle counter clockwise. There exist closed expressions for this integral that can be evaluated using the residue theorem, such as in the case of $\ms=\frac12$ in which one finds Eq.~\eqref{eq:HilbSpaceDimension}, which is derived in Appendix~\ref{sec:S2} by other means.

To obtain the asymptotic behavior for large values of $L$, it is better to express $J$ in terms of the spin density $\sd=\frac{\spin}{\ms L}$ and apply the saddle point approximation. One finds 
\begin{equation}\label{eq:inta}
    n^{\ms}_\spin=i\oint \frac{(z^2-1)z}{2\pi} e^{\psi_\ms(z)L}dz,
\end{equation}
with
\begin{align}
    \psi_\ms(z)=2\,\ms\,j\ln(z)+\ln\left(\tfrac{z^{1+2\ms}-z^{-(1+2\ms)}}{z-z^{-1}}\right)\,.\label{eq:psi-def}
\end{align}
The saddle point approximation then states that the integral in Eq.~\eqref{eq:inta} is approximately given by
\begin{align}
    n^{\ms}_\spin=i\sqrt{\frac{2\pi}{-\psi''_\ms(z_0)L}} \frac{(z^2_0-1)z_0}{2\pi}e^{\psi_\ms(z_0)L}\,,\label{eq:nms-general}
\end{align}
where $z_0$ is the dominating saddle point, such that $\psi'_\ms(z_0)=0$ and $\mathrm{Re}[\psi_\ms(z_0)]$ is maximal. Here, this corresponds to the solution $z_0$ with $\psi'_\ms(z_0)=0$ that is non-negative and real.

For $\ms=\frac12$, the dominating saddle point is $z_0=\sqrt{(1-j)/(1+j)}$, which gives rise to the $\beta^{1/2}(j)$ reported in Eq.~\eqref{eq:AsymptoticHilbSpcDim}, and to the approximation in Eq.~\eqref{eq:AsymptoticHilbFraction}. For $\ms=1$, the dominating saddle point is
\begin{align}
    z_0(j)=\sqrt{\frac{\sqrt{4-3j^2}-j}{2(1+j)}}\,,
\end{align}
which gives rise to the $\beta^{1}(j)$ reported in Eq.~\eqref{eq:beta-frakj=1}. While one can compute $\alpha(j)$ based on Eq.~\eqref{eq:nms-general}, understanding $\beta(j)$ suffices to advance the leading volume-law term of the average entanglement entropy.

\section{Hilbert space dimensions for $\ms=\frac12$}\label{sec:S2}

For the specific case of $\ms=\frac12$, one can find closed-form expressions for the Hilbert space dimensions using combinatorics~\cite{Dim_Spin_Half_1999,Dim_Spin_Half_2005,Cohen2016}. For completeness, next we summarize how this is done. To lighten the notation, since we only discuss the case $\ms=\frac12$, we drop $\ms$ from all the equations in this appendix. 

Once again, we construct the Hilbert space $\mathcal{H}$ as $\vol$ tensor products of the spin-$\frac12$ representation of $\mathrm{SU}(2)$:
\begin{align}
    \underbrace{\tfrac{1}{2}\otimes\tfrac{1}{2}\otimes\dots\otimes\tfrac{1}{2}}_{\vol\text{ times}}\,.
\end{align}
We can use the rule
\begin{align}
    \spin_1\otimes \spin_2=\bigoplus^{\spin_1+\spin_2}_{\spin=|\spin_1-\spin_2|}\spin\,,
\end{align}
to write
\begin{align}
\begin{split}
    (\tfrac{1}{2})^{\otimes 0}&=0\,,\\
    \tfrac{1}{2}&=\tfrac{1}{2}\,,\\
    \tfrac{1}{2}\otimes\tfrac{1}{2}&=0\oplus 1\,,\\
    \tfrac{1}{2}\otimes\tfrac{1}{2}\otimes\tfrac{1}{2}&=\tfrac{1}{2}\oplus \tfrac{1}{2}\oplus \tfrac{3}{2}\,,\\
    \tfrac{1}{2}\otimes\tfrac{1}{2}\otimes\tfrac{1}{2}\otimes\tfrac{1}{2}&=0\oplus 0\oplus 1\oplus 1\oplus 1\oplus 2\,,\\
    &\,\,\,\vdots\\
    (\tfrac{1}{2})^{\otimes \vol}&=\underbrace{\spin_1\oplus \dots \spin_1}_{n_{\spin_1}\text{ times}}\oplus\dots\oplus\underbrace{\spin_k\oplus \dots \spin_k}_{n_{\spin_k}\text{ times}}\,.
\end{split}
\end{align}
The general form of the multiplicities $n_{\spin}$ can be deduced from a generalization of Pascal's triangle, where we cut the triangle at the middle axis (corresponding to $\spin=0$).
\begin{align*}
    \begin{array}{ccccccccc}
     \text{spin } \spin &     & 0 & \tfrac{1}{2} & 1 & \tfrac{3}{2} & 2 & \tfrac{5}{2} & 3\\[2mm]
     \vol=0 && 1 & & & & & &\\
     \vol=1 && & 1 & & & & &\\
     \vol=2 && 1 & & 1 & & & &\\
     \vol=3 &&  & 2& & 1 & & &\\
     \vol=4 && 2 & & 3&  & 1 & &\\
     \vol=5 &&  & 5& & 4 &  & 1 &\\
     \vol=6 && 5 & & 9 &  & 5 &  &1
    \end{array}
\end{align*}
The entries of the triangle represent the multiplicities $n_\spin$, where $\spin$ consists of positive half-integers for odd $\vol$ and non-negative integers for even $\vol$.

One can find a closed formula for $n_\spin$ as a function of $\vol$ by identifying the process with a random walk on non-negative integers (representing $2\spin$) starting at $0$, where we jump from $0$ to $1$ with probability $1$, while for all other integers $2\spin$, we jump either to $2\spin-1$ or $2\spin+1$ with probability $\frac12$ each. The number of paths leading to integer $2\spin$ after $\vol$ steps can then be calculated using Bertrand's ballot theorem~\cite{feller1968introduction, rensburg-book} (in the variant where ties are allowed). In this context, the random walk is yet again re-interpreted as counting ballots for two candidates with total votes $p$ for the candidate 1 and $q<p$ votes for candidate 2.  Bertrand's ballot theorem (ties allowed) then states that the number of ways the votes can be counted (one after each other), such that candidate 1 is never behind candidate 2 is given by
\begin{align}
    \binom{p+q}{q}-\binom{p+q}{q-1}=\frac{p+1-q}{p+1}\binom{p+q}{q}\,.
\end{align}
In our case, we have $p+q=\vol$ (total votes) and $2\spin=p-q$ ($p$ represents right-steps and $q$ represents left-steps). With this, we find
\begin{align}
    n_\spin=n_\spin(\vol)=\frac{2(1+2\spin)}{2+\vol+2\spin}\binom{\vol}{\frac{\vol}{2}-\spin}\,.\label{eq:mS}
\end{align}
Note that $\vol/2-\spin$ is always an integer, as $\spin$ is a half-integer whenever $\vol$ is odd.

Based on this calculation, we can determine the dimensions of the Hilbert spaces with fixed total spin $\spin$, fixed spin $\spin_z$, and fixing both $\spin$ and $\spin_z$. The corresponding dimensions are then given by
\begin{align}
    \dim(\spin_z)&=\binom{\vol}{\frac{\vol}{2}+\spin_z}\,,\label{eq:dim1}\\
    \dim(\spin)&=\frac{2(1+2\spin)^2}{2+\vol+2\spin}\binom{\vol}{\frac{\vol}{2}-\spin}\,,\\
    \dim(\spin_z,\spin)&=\begin{cases}
    \frac{2(1+2\spin)}{2+\vol+2\spin}\binom{\vol}{\frac{\vol}{2}-\spin} & \spin\geq |\spin_z|\\
    0 & \text{otherwise}
    \end{cases}\label{eq:dim3}\,,
\end{align}
and we see that, as long as $\spin\geq |\spin_z|$, the dimension of the Hilbert space for fixed $(\spin,\spin_z)$ is independent of $\spin_z$. Hence, the Hilbert space dimension of a sector with fixed $\spin$ within the $\spin_z=0$ subspace is $\dim(\spin_z=0,\spin)=n_J$. 

\section{Maximally chaotic regime}\label{sec:S1}

In order to reduce finite-size effects in the comparison between the average entanglement entropy of highly excited eigenstates of a one-dimensional quantum-chaotic (nonintegrable) Hamiltonian and random pure states with spin $\ms=\frac12$, following the recent discussion in Ref.~\cite{kliczkowski2023average} we set the Hamiltonian parameter $\cnn=3$ [see Eq.~\eqref{eq:Hamiltonian Spin 1/2} in the main text] to be in the maximally chaotic regime. By maximally chaotic regime it is meant that, for the system sizes that one can study using exact diagonalization, sensitive probes of quantum chaos return results that are closest to the random matrix theory predictions.

To locate the maximally chaotic regime, we use translational invariance to diagonalize the Hamiltonian in the zero magnetization sector ($\spin_z=0$). Translational invariance allows us to block diagonalize the Hamiltonian within sectors with total quasimomentum $k=2n\pi/\vol,\ n\in [0,\vol/2]$. We consider chains with $\vol=20$ and $22$, and focus on the ``complex'' sectors with $n\in [1,\vol/2-1]$. Those sectors lack the reflection symmetry present in the ``real'' $k=0$ and $\pi$ sectors, and suffer from smaller finite-size effects~\cite{leblond_entanglement_2019, kliczkowski2023average}. We select the central 100 eigenstates with $\spin=0,1,$ and 2 in each of the complex sectors, in each eigenstate we compute the two quantum chaos indicators mentioned below, and then average the results over all the eigenstates with a given value of $\spin$.

\begin{figure}[!t] 
    \centering
    \includegraphics[width=0.98\columnwidth]{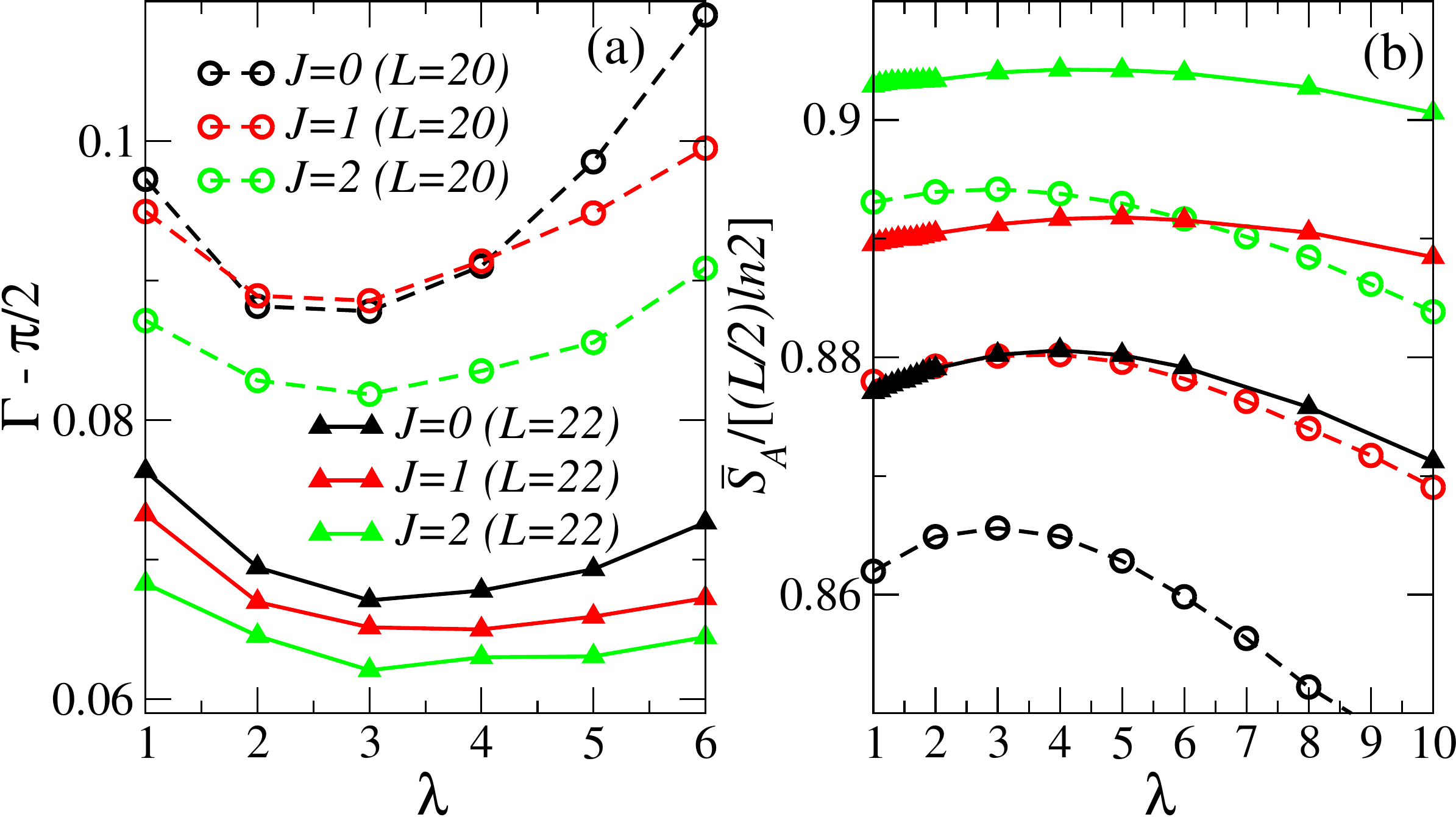}
    \caption{{\it Maximally chaotic regime}. Results for: (a) $\Gamma-\pi/2$ and (b) $\bar \ent_A$ at subsystem fraction $f=\frac12$ plotted as functions of $\cnn$ for $\spin=0,\,1,$ and 2 in chains with $\vol=20$ and 22, for the microscopic spin $\ms=\frac12$.}
    \label{MaximalChaos}
\end{figure}

The two quantities that we compute in each eigenstate are the ``Gaussianity'' and the entanglement entropy at $f=\vol_A/\vol=\frac12$~\cite{kliczkowski2023average}. The Gaussianity is defined as
\begin{equation}\label{eq:gauss}
    \Gamma_n=\frac{\overline{|x_\alpha^{(n)}|^{2}}}{\overline{|x_\alpha^{(n)}|}^{\,2}}\,,
\end{equation}
where $x_\alpha^{(n)}=\text{Re}[C_\alpha^{(n)}]$, $C_\alpha^{(n)}$ being the coefficient of total quasimomentum eigenstate $\ket{k_\alpha}$ (with the appropriate $Z_2$ eigenvalue within the $\spin_z=0$ sector) in the energy eigenstate $\ket{E_n}$, $\ket{E_n}=\sum_{\alpha}C_\alpha^{(n)}\ket{k_\alpha}$. (We obtain similar results, not shown, using $\text{Im}[C_\alpha^{(n)}]$.) The averages in Eq.~\eqref{eq:gauss} are computed over $i$, and then we further average $\Gamma_n$ over all eigenstates with a given $\spin$ to obtain $\Gamma=\bar\Gamma_n$ reported in Fig.~\ref{MaximalChaos}(a). Since the eigenstates of random matrices are random unit vectors with normally distributed coefficients, the random matrix prediction for $\Gamma$ is $\Gamma_\text{RM}=\pi/2$~\cite{leblond_entanglement_2019}. 

Figure~\ref{MaximalChaos} shows our results for $\Gamma-\pi/2$ [Fig.~\ref{MaximalChaos}(a)] and for $\bar \ent_A$ [Fig.~\ref{MaximalChaos}(b)] as functions of $\cnn$. The results in Fig.~\ref{MaximalChaos}(a) show that $\Gamma$ is closest to the random matrix theory prediction for the three values of $J$ considered for $\vol=20$ and 22, about $\cnn = 3$. For the average entanglement entropy in Fig.~\ref{MaximalChaos}(b), we find that the maximum occurs between $\cnn=2$ and $6$ depending on the value of $\vol$ and $\spin$. Given those results, we selected $\cnn=3$ in the maximally chaotic regime to carry out the finite-size scaling analyses reported in the main text.

\section{Asymptotics of SD$_{\text{2}}$} \label{sec:S3}

We extract the large-$\vol$ asymptotics of Eq.~\eqref{eq:SA-SD_B} as explained below. The general method is similar to the one explained in detail in Ref.~\cite{bianchi2022volume} to compute $\braket{S_A}_{N}$.

First, we compute the asymptotic $d_{J_A}$ in terms of $\vol$, $f$, and the subsystem spin density $j_A=2J_A/L$, to find
\begin{eqnarray}
    && d_{J_A} \! = \! \frac{\alpha}{\vol}\exp\!\left\{\left[f\, s_A \!\! \left(\frac{j_A}{f}\right)+(1-f)\, s_A\!\! \left(\frac{j-j_A}{1-f}\right)\right]\! L\right\}, \nonumber\\ \\ 
    &&\alpha=\frac{8j_A(j-j_A)}{\pi(1-f+j-j_A)(f+j_A)}\nonumber \\&& \hspace{0.7cm}\times \sqrt{\frac{(1-f)f}{(f^2-j_A^2)(1-f-j+j_A)(1-f+j-j_A)}}\nonumber \\&&\hspace{0.7cm} + \, o(1) \,,
\end{eqnarray}
where $s_A(\cdot)$ is defined in Eq.~\eqref{eq:CoefficientJ=O(L)}.

Second, we approximate $\rho(j_A)=\frac{\vol}{2}\frac{d_{J_A}}{d}$ by a Gaussian using a saddle-point approximation around the mean $\bar{j}_A=jf$. We find that the variance is given by $\sigma^2=(1-j^2)f(1-f)/L+O(1/L^2)$. We Taylor expand the exponent of $d_{J_A}$ up to cubic order around $\bar{j}_A$ and then expand the exponential up to linear order to find
\begin{eqnarray}
    \rho(j_A)&=&\frac{1}{\sqrt{2\pi\sigma^2}}\exp\left[-\frac{(j_A-\bar{j}_A)^2}{2\sigma^2}\right]\nonumber\\&&\times\left[1+\sum_{\ell=1,3}\alpha_\ell(j_A-\bar{j}_A)^\ell+o(1)\right]\!,
\end{eqnarray}
where $1/(\sqrt{2\pi\sigma^2})$ normalizes the Gaussian, and the expansion coefficients $\alpha_\ell$ (note that the quadratic order is absorbed in the definition of the Gaussian) are given by
\begin{align}
    \alpha_1&=\frac{(1-2f)(1-j+j^2)}{(1-f)fj(1-j^2)}+o(1)\,,\\
    \alpha_3&=\frac{(1-2f)j L}{3(1-f)^2f^2(1-j^2)^2}+O(1)\,,
\end{align}
where the $O(1)$ term in $\alpha_3$ will only contribute towards an $o(1)$ term in the final result.

Third, we use that the CG coefficients $c_m(J,J_A,J-J_A)$ follow a normal distribution with zero mean and variance $\sigma^2_m=\spin_A(\spin-\spin_A)/(2 \spin)=j_A(j-j_A)L/(4j)$ for large $L$. The entropy of the normal distribution is
\begin{align}
    \ent_{\mathrm{CG}}(\spin_A)=\ln(\sqrt{2\pi e}\,\sigma_m)+o(1)\,.
\end{align}

Fourth, we replace the sum in Eq.~\eqref{eq:SA-SD_B} over $\spin_A$ by an integral over the subsystem spin density $j_A$, \ie $\sum_{J_A}\to \frac{\vol}{2}\int dj_A$ and split the summand, now an integrand, into the product of $\rho(j_A)$, which includes the factor $L/2$, and
\begin{eqnarray}
  \hspace{-0.35cm}  &&\varphi(j_A)=L\left[s_A(j)-(1-f)s_A\!\!\left(\frac{j-j_A}{1-f}\right)\right]\\
  \hspace{-0.35cm}  &&+\ln\!\!\left[\frac{2j^2(1-f+j-j_A)\sqrt{(1-f)(1-j^2)(1-\frac{(j-j_A)^2}{(1-f)^2})}}{(1-j)(1+j)^3(j-j_A)}\right]\nonumber\\
\hspace{-0.35cm}   &&+\frac{1}{2}\ln\left[\frac{\pi e j_A(j-j_A)L}{2j}\right]+o(1)\,.\nonumber
\end{eqnarray}
There is an important subtlety, namely, $\varphi(j)$ is non-analytical at $j_{\mathrm{crit}}=\frac{2}{\vol}\spin_{\mathrm{crit}}$ (defined as the point where $n^A_{\spin_{\mathrm{crit}}} = n^B_{\spin_{\mathrm{crit}}}$), such that for $j_A\geq j_{\mathrm{crit}}$, we need to replace $f\to 1-f$ and $j_A\to j-j_A$.

Fifth and finally, we carry out the integration by expanding $\varphi(j_A)$ up to quadratic order in $(j_A-\bar{j}_A)$ to find Eq.~\eqref{eq:EntanglementJ=O(L)} for $0<j<1$. Note that the $\sqrt{\vol}$ term with the Kronecker delta at $f=\frac12$ stems from the alignment of the center of the Gaussian $\bar{j}_A=f j$ and $j_{\mathrm{crit}}$, such that the Taylor expansion of $\varphi(j_A)$ is different for $j_A\leq \bar{j}_A$ and $j_A\geq \bar{j}_A$.

\section{Complex vs real random coefficients}\label{sec:randcoeff}

\begin{figure}[!t]
    \centering
    \includegraphics[width=0.98\columnwidth]{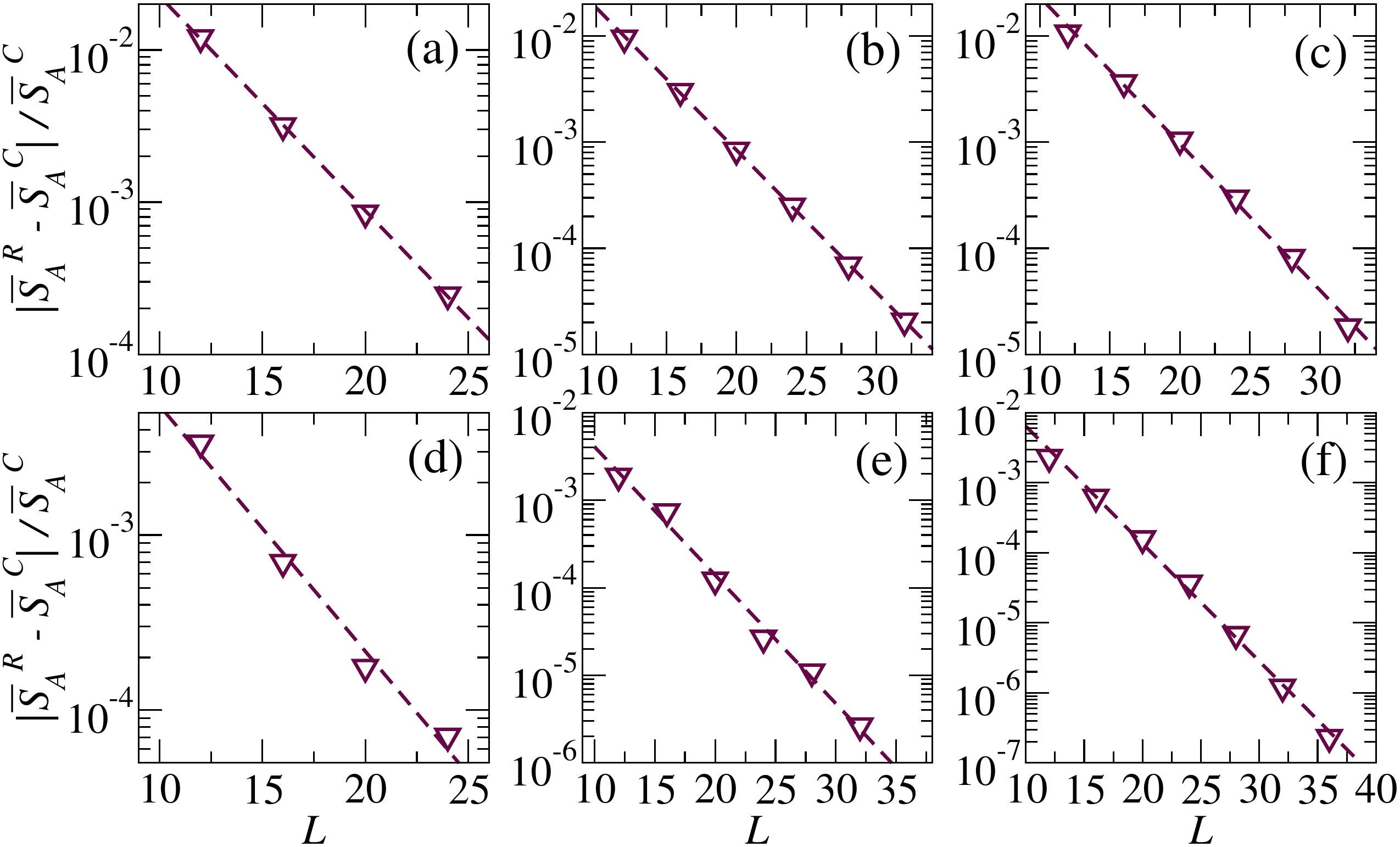}
    \caption{{\it Gaussian distributed real vs complex coefficients.} Relative difference between $\bar \ent_A$ obtained numerically by sampling real ($\bar \ent^R_A$) vs complex ($\bar \ent^C_A$) coefficients for $\spin=L/4$. The columns correspond to the results for random states [(a) $f=\frac12$ and (d) $f=\frac14$], SD$_\text{1}$ [(b) $f=\frac12$ and (e) $f=\frac14$], and SD$_\text{2}$ [(c) $f=\frac12$ and (f) $f=\frac14$]. In all cases the results are consistent with an $e^{-a\vol}$ decay with the number of lattice sites $\vol$, as indicated by the dashed lines.}\label{fig:Real-ComplexJ=L/4}
\end{figure}

To compute all the numerically obtained average entanglement entropies $\bar \ent_A$ reported in the main text: for random pure states, SD$_{\text{1}}$, and SD$_{\text{2}}$, we use Gaussian distributed {\it real} coefficients, as opposed to the Gaussian distributed {\it complex} coefficients implicit in the Haar-random averages carried out in our analytical calculations. Real coefficients are used in the numerical calculations to reduce the computation time. As shown in Fig.~\ref{fig:Real-ComplexJ=L/4}, the relative differences between the results obtained using real and complex coefficients decreases exponentially with increasing $L$, and it is very small for the systems sizes considered in our study.

All the results reported for random pure states were obtained by averaging over at least 1000 random states for $\vol\leq 20$ and over at least 100 random states for $\vol>20$. All the results reported for the SD$_\text{1}$ and SD$_\text{2}$ approximations were obtained by averaging over at least 1000 random states for $L\leq30$, and over at least 100 random states for $L>30$.\\

\newpage

\bibliography{references}

\end{document}